\RequirePackage{fix-cm}
\documentclass[smallcondensed]{svjour3} 

\usepackage{todonotes}
\usepackage{natbib}
\PassOptionsToPackage{table,xcdraw}{xcolor}
\definecolor{main}{HTML}{CFCFCF}    % Setting main color
\definecolor{sub}{HTML}{CFCFCF}     % Setting sub color

\usepackage{multirow}

\usepackage{blindtext} % For generating placeholder text
\usepackage{url}
\usepackage[unicode]{hyperref}

\usepackage{fontawesome}

\usepackage[many]{tcolorbox} 
\tcbset{
    sharp corners,
    colback = white,
    before skip = 0.5cm,    % Extra space before the box
    after skip = 0.7cm      % Extra space after the box
}

\newtcolorbox{boxCNoTitle}{
    top=10pt,
    rounded corners,
    coltitle=black,
    colframe=gray,
    colbacktitle=sub,
    colback = sub,
    enhanced,
    center,
    boxrule=0.5pt
}
\newtcolorbox{boxC}[2][]{aibox,title=#2,#1}
\tcbset{
  aibox/.style={
    top=10pt,
    boxrule=0.5pt,
    rounded corners,
    coltitle=black,
    colframe=gray,
    colbacktitle=sub,
    colback = sub,
    enhanced,
    center,
    attach boxed title to top left={yshift=-0.1in,xshift=0.15in},
    boxed title style={boxrule=0.5pt,colframe=gray,},
  }
}

\newcounter{keyTakeAwaysCounter}
\newenvironment{keyTakeAways}[1][Key Take Away]
    {
        \addtocounter{keyTakeAwaysCounter}{1}
        \begin{boxC}{\faLightbulbO ~ \thekeyTakeAwaysCounter. #1}
        
    }{
        \end{boxC}
    }

\newcounter{keyRQAnswerCounter}
\newcounter{keyLimitationsCounter}
\newenvironment{keyLimitations}[1][Key Limitations]
    {
        \addtocounter{keyLimitationsCounter}{1}
        \begin{boxC}{\faWarning ~ \thekeyLimitationsCounter. #1}
        
    }{
        \end{boxC}
    }
\usepackage{amssymb}  % Mathematical symbols
\usepackage{lineno}

\usepackage{listings}

\lstdefinelanguage{json}{
    basicstyle=\ttfamily\small,          % Font style
    numbers=none,                        % Disable line numbers
    frame=lines,                         % Add a frame around the code
    breaklines=true,                     % Enable word wrapping
    breakatwhitespace=false,             % Allow breaks anywhere
    postbreak=\mbox{\textcolor{red}{$\hookrightarrow$}}, % Symbol for line breaks
    showstringspaces=false,              % Don't show spaces in strings
    literate=%
          *{0}{{{\color{blue}0}}}{1}%
         {1}{{{\color{blue}1}}}{1}%
         {2}{{{\color{blue}2}}}{1}%
         {3}{{{\color{blue}3}}}{1}%
         {4}{{{\color{blue}4}}}{1}%
         {5}{{{\color{blue}5}}}{1}%
         {6}{{{\color{blue}6}}}{1}%
         {7}{{{\color{blue}7}}}{1}%
         {8}{{{\color{blue}8}}}{1}%
         {9}{{{\color{blue}9}}}{1}%
         {:}{{{\color{red}:}}}{1}%
         {\{}{{{\color{gray}\{}}}{1}%
         {\}}{{{\color{gray}\}}}}{1}%
         {[}{{{\color{gray}[}}}{1}%
         {]}{{{\color{gray}]}}}{1}%
         {"}{{{\color{darkgreen}"}}}{1}%
         {\"[^\"]*\"}{{{\color{darkgreen}\lst@string}}}{0},  % Entire string in dark green
}

\definecolor{darkgreen}{rgb}{0,0.5,0}

\newcommand{\expertNo}{33 }

\begin{document}

%\begin{frontmatter}

% Short author
%\shortauthors{Esposito et al.}

% Authors
\author{Matteo Esposito         \and
        Mikel Robredo           \and
        Murali Sridharan        \and
        Guilherme Horta Travassos \and
        Rafael Pe\~naloza       \and
        Valentina Lenarduzzi}

% Affiliations
% Affiliations
\institute{Matteo Esposito \at
              University of Oulu, Oulu, Finland \\
              \email{matteo.esposito@oulu.fi} - ORCID: 0000-0002-8451-3668
           \and
           Mikel Robredo \at
              University of Oulu, Oulu, Finland \\
            \email{mikel.robredo@oulu.fi} - ORCID: 0009-0001-9870-1504
           \and
           Murali Sridharan \at
              University of Oulu, Oulu, Finland \\
            \email{murali.sridharan@oulu.fi} - ORCID: 0000-0002-5212-588X
           \and
           Guilherme Horta Travassos \at
              Federal University of Rio de Janeiro, Rio de Janeiro, Brazil \\
              CNPq Researcher and FAPERJ CNE\\
              \email{ght@cos.ufrj.br} - ORCID: 0000-0002-4258-0424
           \and
           Rafael Pe\~naloza \at
              University of Milano-Bicocca, Milano, Italy \\
              \email{rafael.penalozanyssen@unimib.it} - ORCID: 0000-0002-2693-5790
           \and
           Valentina Lenarduzzi \at
              University of Oulu, Oulu, Finland \\
              \email{valentina.lenarduzzi@oulu.fi} - ORCID: 0000-0003-0511-5133
}

% Corresponding author
%\cortext[1]{Corresponding author: Matteo Esposito (matteo.esposito@oulu.fi)}

\date{Received: date / Accepted: date}

\title{A Call for Critically Rethinking and Reforming\texorpdfstring{ \\}{ }Data Analysis in Empirical Software Engineering}
% Short title
%\shorttitle{A Call for Critically Rethinking and Reforming Data Analysis in Empirical Software Engineering}

\titlerunning{A Call for Critically Rethinking and Reforming Data Analysis in ESE}
\date{Received: date / Accepted: date}
% The correct dates will be entered by the editor

\maketitle

\begin{abstract}
\noindent \textit{Context.}  Empirical Software Engineering (ESE) is the core of innovation in the broader SE fields via its qualitative and quantitative empirical studies. Albeit empirical methodologies have driven the advances in SE practices and research, concerns over their correct application have sparked as early as 2006 in the Dagstuhl seminar on SE. So far, advancements would be expected since some previous studies have investigated it in specific software engineering areas, highlighting the misconceptions about using inadequate statistics strategies to analyze data. However, it is unclear how it has evolved since then and how it could influence confidence in decision-making based on the results of primary studies. 

\noindent \textit{Objective.} To analyze three decades of SE research and look for eventual mistakes in choosing adequate statistics strategies to support the data analysis, i.e., flawed statistical techniques for analyzing data reported in the technical literature. Besides, it will evaluate current experts' accuracy in detecting and addressing these issues.

\noindent \textit{Methods.}  To perform a literature survey (pilot study) in the technical literature to collect a large sample of empirical studies, i.e., circa 27k works, reporting data analysis, and filtering them using LLM to identify adequate (``good'') and inadequate (``bad'') statistical methodological patterns. To randomly select 30 primary studies (15 good and 15 bad)  and conduct a focus group-based workshop with 33 Empirical Software Engineering experts to observe their capabilities in identifying and remediating those methodological statistics issues.

\noindent \textit{Results.} Our findings reveal widespread statistics issues within the published primary studies regarding empirical software engineering studies. Besides, the performance of the experts in identifying eventual mistakes in the use of statistical methods, models, and tests in these primary sources raises serious concerns about the general ability of empirical software engineering researchers to spot them and suggest proper adjustments. 

\noindent \textit{Conclusions.} Despite our study's eventual limitations, its results shed light on recurring issues from promoting information copy-and-paste from past authors' works and the continuous publication of inadequate approaches that promote dubious results and jeopardize the spread of the correct statistical strategies among researchers.  Besides, it justifies further investigation into empirical rigor in software engineering to expose these recurring issues and establish a framework for reassessing our field's foundation of statistical methodology application. Therefore, this work calls for critically rethinking and reforming data analysis in empirical software engineering, paving the way for our work soon.
\end{abstract}
%\linenumbers


%\end{frontmatter}

% \linenumbers

\section{Introduction}
\label{sec:intro}

The term ``Software Engineering'' first emerged in the mid-20th century, with early usages including a 1965 letter from ACM president Anthony Oettinger~\citep{meyer2013origin,tedre2014science} and lectures by Douglas T. Ross at MIT in the 1950s~\citep{mahoney1990roots}. These pioneers recognized the need for a disciplined approach to software development. Margaret H. Hamilton, when working on the Apollo Guidance Computer, coined the term ``software engineering'' (SE) to denote the rigor and importance of the practices involved~\citep{rayl2008nasa,hamilton2018icse}. SE was popularized during the 1968 NATO Software Engineering Conference, which addressed the ``software crisis'', i.e., the challenges of developing complex software systems that were over budget, behind schedule, and low-quality~\citep{hackreactor_history_2020}. 
In response to these challenges, researchers recognized the need for a more scientific approach to software development. This led to adopting empirical methods, such as controlled experiments, case studies, and surveys, to systematically evaluate software processes and tools. Victor Basili, an empirical software engineer pioneer, advocates for using empirical studies to build a science of computer science~\citep{Basili1994}. 

Over the decades, Empirical Software Engineering (ESE) has evolved to encompass various research methods, including quantitative and qualitative approaches \cite{Felderer2020}. This evolution reflects a growing recognition of the importance of empirical evidence in understanding and improving software engineering practices.
However, early warnings regarding the uncritical adoption of empirical methodologies (EM), including the tendency to replicate methodological issues through widespread copy-paste practices, were raised as early as 2006 during the Dagstuhl seminar on Software Engineering \citep{DBLP:conf/dagstuhl/2006esei}.  The legacy of such a seminar was considered in \cite{reyes2018statistical}, where they conducted an empirical study to assess the extent of statistical errors in software engineering experiments. The authors compiled a list of frequent statistical mistakes from various experimental disciplines and systematically reviewed all papers published at ICSE. Finally,  in \cite{vegas2023pitfalls}, they analyzed pitfalls in experiments with Deep Neural Networks (DNNs) in SE. 
\cite{Vegas2024} warned the community that specific methodological issues could push SE to the situation faced by psychology, i.e., a reproducibility crisis.

Stemming from the daunting need to verify and evaluate the degree to which our field falls into a methodological crisis, we investigate using methodological statistical strategies in ESE studies, distinguishing between adequate patterns that adhere to best practices and inadequate patterns. On the other hand, we evaluate the experts' inadequate methodological pattern detection and remediation capabilities. 
The study draws on a rich corpus of ESE literature spanning the last 30 years (1994–2023). These studies encompass various domains and methodologies, reflecting the evolution of empirical data analysis practices in ESE. Therefore, we split our study into two sub-studies.

To evaluate the actual trends over the year, we devised a \textbf{pilot study} that surveys a portion of the state of the art. Based on 60.381 original studies, our pilot findings alarmingly highlighted many works with methodological issues published in the last 30 years. Stemming from these findings, we conducted a \textbf{workshop} with \expertNo ESE specialist to evaluate their abilities in determining statics misuses when analyzing empirical data. 

Our findings reveal widespread methodological issues within the ESE field, including misidentifying statistical methods, models, and tests. This raises serious concerns about experts' performance in spotting them and indicates risks of spreading false evidence regarding software phenomena. Shortly, we aim to expose these recurring issues extensively and establish a framework for reassessing the very foundation of research in our field.  Therefore, this work calls for critically rethinking and reforming data analysis in ESE, paving the way for our future work.

\section*{Paper Structure}   
Our study is composed of two sub-studies, namely a pilot study investigating the usage of methodological patterns in ESE studies, distinguishing between adequate (``good'')
patterns that adhere to best practices and inadequate (``bad")
pattern and a workshop evaluating the experts’ inadequate methodological pattern detection and remediation capabilities. This section briefly presents the structure of our study.

\textbf{Study Overview}. Section~\ref{sec:BG} presents the background, Section~\ref{sec:RW} presents the related work, and Section~\ref{sec:methods} describes the overall study design. 

\textbf{Pilot Study}. Section~\ref{sec:pilot} presents the pilot study. Specifically, 
Section~\ref{sec:pilot:methodology} presents the pilot study design, Section~\ref{sec:pilot:results} presents the obtained results, Section~\ref{sec:pilot:discussions} discusses them, and Section~\ref{sec:pilot:threats} highlights the threats to the validity of our pilot study.  

\textbf{Workshop}. Section~\ref{sec:workshop} presents the workshop. Specifically, 
Section~\ref{sec:workshop:methodology} presents the workshop study design, Section~\ref{sec:workshop:results} presents the results, Section~\ref{sec:workshop:discussions} discusses them, and Section~\ref{sec:workshop:threats} highlights the threats to the validity of our workshop.  

\textbf{Conclusions}.  Section~\ref{sec:insightsAndLimitations} discusses the key insights of our overall study and the main limitations.  
Finally, Section~\ref{sec:conclusions} concludes.

%Leveraging recent peer review experience and following in the footsteps of the Dagstuhl above seminar, we started questioning whether some methodological issues slipped into published work.

%In this work we conducted a pilot study with \expertNo renown experts from the ESE community to investigate ESE expert proficiency in identifying methodological issues from currently published papers. The contribution of our study is two folded: (i) on the one hand we survey the expert community in ESE for detecting the methodological issues: (ii) on the other hand we spark the discussion on the validity of past studies paving the way for un upcoming study surveying the last 30 years of ESE methodologies.

\section{Background}
\label{sec:BG}
This section presents the background concept that lies as the foundation of our study.

\subsection{Empirical Studies} 
\label{subsec:es}
In SE, \citet{wholin2012experimentation} identified various types of qualitative and quantitative empirical studies that researchers use to explore and enhance practices in the field. These methods offer distinct ways to investigate complex phenomena and tackle research questions, each with strengths and trade-offs.

\textbf{Case studies} focus on in-depth explorations of specific, real-world contexts, such as a particular software development project or team. They excel at uncovering the “how” and “why” behind outcomes, providing rich, detailed insights \citep{lenarduzzi2017prioritizing}. For instance, a case study might examine how a team adopts agile practices in a fast-paced environment. While these studies reveal valuable nuances, their findings often struggle to extend to broader contexts.

Authors of \textbf{controlled experiments} manipulate independent variables and observe their impact on dependent variables in a controlled setting. This approach is ideal for testing hypotheses and identifying causal relationships \citep{Basili1994}. For instance, a study might evaluate many static analysis security testing tools to see which helps developers detect vulnerabilities more reliably~\citep{esposito2024extensive}. The controlled environment ensures precision but often sacrifices real-world relevance.

Researchers frequently turn to \textbf{surveys} for a broader view. These studies collect data from large groups of participants using questionnaires or interviews. Surveys are ideal for spotting trends, gauging opinions, or understanding behaviors across diverse populations \citep{torchiano2017lessons}. However, their breadth usually comes at the cost of the depth offered by case studies or the experimental rigor of controlled studies.

\textbf{Action research} takes an interactive and iterative path, with researchers actively engaging in problem-solving within the study environment. This approach blends practical problem-solving with knowledge generation, making it invaluable for connecting theory to real-world practice \citep{staron2020action}. For example, researchers might collaborate with a software team to improve their testing process while simultaneously studying the impact of those changes.

\textbf{Ethnographies} dive deeply into cultural and social dynamics by immersing researchers in the daily lives of software engineers or teams \citep{silva2013using}. This qualitative method uncovers subtle, meaningful insights by observing interactions, workflows, and collaboration patterns. While it requires significant time and effort, the rewards are profound, revealing aspects of SE that might otherwise remain hidden.

Each method offers unique advantages, and researchers carefully select the one that best aligns with their goals. Whether the aim is to dissect a specific situation, validate a hypothesis, or explore industry-wide trends, these approaches collectively form a comprehensive toolkit for advancing knowledge and practice in SE.

\subsection{Data Analysis Taxonomy}
\label{subsec:dat}
Data analysis in SE is a vast topic, and it would be beyond the scope of the present paper to comprehensively map all the strategies adopted by past and current authors. Nonetheless, a taxonomy is needed to enable our investigation.
%To enable our investigation and remove any subjective biases to discuss specific methodological errors, we leveraged a systematic taxonomy of keywords to capture and collect specific methodological choices from a given study. 
Therefore, we compile our data analysis taxonomy leveraging the work of \citet{livingstone2009practical} and \citet{meloun2011statistical}, which are the widest adopted statistical manuals, based on suggestions of colleagues and experts in the statistic domain. In other words, \citet{livingstone2009practical,meloun2011statistical}'s guidelines are for statisticians what \citet{DBLP:conf/icse/KitchenhamDJ04,  DBLP:journals/ese/RunesonH09, wholin2012experimentation}'s guidelines are for our community. Moreover, for each category and item of the categories, we double-checked the original author paper in which the technique and test were defined. Therefore, we avoided personal interpretation of any collectors, surveyors, or users of such techniques referencing only the original author's interpretation of their work, e.g., Spearman's interpretation of its $\rho$. Therefore, our taxonomy includes the following categories and items: 
\begin{itemize}
    \item \textbf{Variable type}~\citep{kaliyadan2019types}. It refers to the categorization of variables (e.g., nominal, ordinal, interval, or ratio) and their roles (e.g., dependent, independent) in analysis;
    \item \textbf{Distributional check}~ \citep{meloun2011statistical}.  It involves assessing whether data follows a specific distribution, such as normality, using tests (e.g., Shapiro-Wilk) or visual methods (e.g., Q-Q plots).
    \item \textbf{Preprocessing}~\citep{livingstone2009practical}.  It covers data preparation steps like cleaning, normalization, transformation, and handling missing values to make data suitable for analysis.
    \item \textbf{Models}~\citep{agresti2015foundations}. They encompass the mathematical, statistical, or AI-based frameworks used for analysis, such as regression, classification, and clustering algorithms or machine learning and neural networks.
    \item \textbf{Performance evaluation}~\citep{botchkarev2018performance}.  It focuses on metrics and techniques to assess the effectiveness of models, such as accuracy, precision, recall, or error rates.
    \item \textbf{Hypothesis testing}~\citep{statsandrWhatStatistical}. It involves statistical tests to determine whether there is enough evidence to reject a null hypothesis, such as t-tests, ANOVA, or chi-squared tests;
    \item \textbf{Post-hoc} correction~\citep{armstrong2010post}. It refers to adjustments made to account for multiple comparisons in hypothesis testing, reducing the risk of Type I errors, using methods like Bonferroni or Holm corrections.

\end{itemize}
We provide the full taxonomy in the replication package (see data availability statement).   

\subsection{Methodological Statistical Pattern}
\label{sec:Pattern}
We define a methodological pattern as a \textbf{structured pathway} or \textbf{sequence of steps} when applying specific methods to analyze data, test hypotheses, or evaluate results. These patterns guide the researcher in choosing appropriate tools, techniques, and tests to ensure valid and reliable outcomes. A pattern can be seen as a roadmap where each decision or step is drawn from our taxonomy, ultimately influencing the quality of the analysis. We defined as:

\begin{itemize}
    \item A\textbf{ good (or adequate) methodological pattern} is a pathway that adheres to methodological guidelines, validates assumptions at every stage, and uses appropriate techniques based on data characteristics.
    \item A \textbf{bad (or inadequate) methodological pattern }ignores key assumptions, skips critical validation steps, or misuses statistical techniques.
\end{itemize}

\begin{figure}
    \centering
    \includegraphics[width=0.8\linewidth]{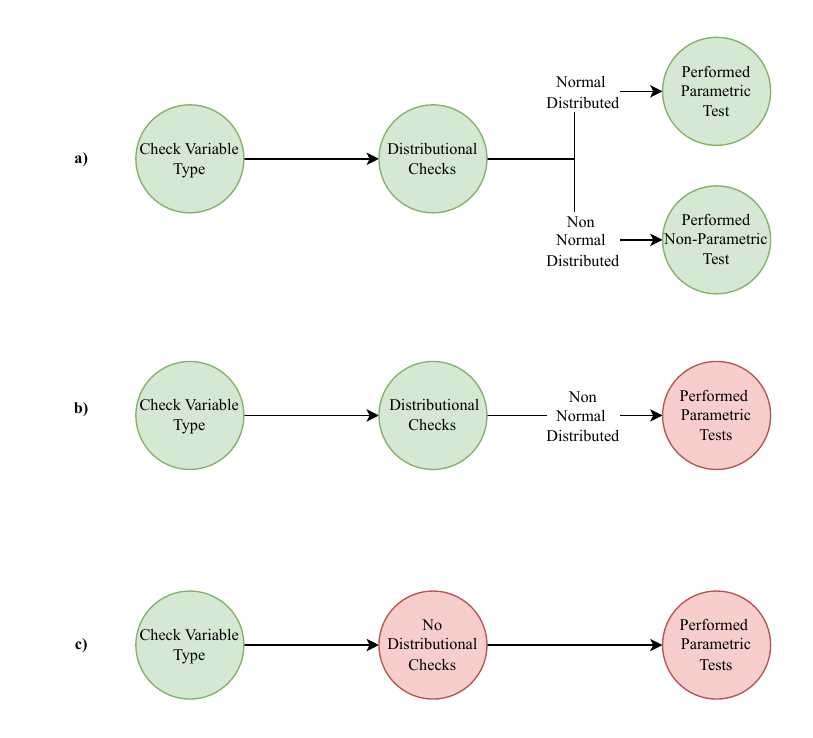}
    \caption{Methodological Patterns Example}
    \label{fig:patternEx}
\end{figure}

For instance, Figure \ref{fig:patternEx} presents three possible patterns:  \textbf{one good} (case \textbf{a}) and \textbf{two bad} (cases \textbf{b, c}). More specifically, for case \textbf{b}, we note that a distributional check was carried out; it resulted in data that was not normally distributed, but a parametric test was performed. In the same vein, \textbf{case c} did not perform a distributional check but performed a parametric test.

\section{Related Works}
\label{sec:RW}
%This section presents related works to our study.
Our investigation stemmed from three main related works upon which our point of view was set. In late 2006, during the Dagstuhl seminar on SE, sparks of methodological derailment were starting to be evident. According to \cite{DBLP:conf/dagstuhl/2006esei}, ESE faced several challenges, especially in data analysis. One of the major issues was the absence of standardized methods for aggregating and validating results across studies. Without systematic methods for meta-analysis, researchers cannot effectively corroborate findings and identify evidence gaps. Current methods, such as vote counting, are usually biased, especially in cases where the sample size is small or the effect size is minimal, Briand 2007. This limitation signifies the need to develop \textbf{meta-analysis methods} since those borrowed from other domains, such as medicine, have not been able to solve specific complexities of software engineering experiments \citep{Oivo2007}.

\textbf{Data quality} is another critical challenge. Researchers often have incomplete or noisy data, especially when working with software repositories. For instance, missing values or default entries can make certain attributes unreliable. These all pose serious challenges in requiring arduous validation procedures for the information to reflect correctly what it purports to reflect \citep{Sjoberg2007}.

In the case of survey studies, \textbf{recollection bias}, where faulty memories distort information, and ``suspicion of exposure'' bias, wherein previous knowledge of an outcome skews the analysis of the information, frequently leads to flawed results in retrospective studies of data \citep{Kitchenham2007}.
A further limitation stems from the poor generalisability of the empirical findings. Most studies concern specific aspects of the software lifecycle, such as coding, and forget entirely earlier stages like requirement engineering and design \citep{Hofer2007}. Moreover, most of the controlled experiments are conducted \textbf{in academic or artificial environments }with students as subjects, which can hardly \textbf{represent} the complexity and dynamics of an \textbf{industrial environment}. This imbalance makes it difficult to apply findings broadly or to gain insights into their practical implications \citep{Hofer2007}.

Reporting practices also fall short, making it hard for researchers and practitioners to locate and interpret experimental results. The lack of standardized reporting guidelines exacerbates this problem, leaving critical information fragmented and inaccessible. Tailoring reports for diverse audiences, i.e., ranging from researchers to industry professionals, remains an unmet need in the field \citep{Vegas2007}.
The legacy of such a seminar was considered in \cite{reyes2018statistical}, where they conducted an empirical study to assess the extent of statistical errors in software engineering experiments. The authors compiled a list of frequent statistical mistakes from various experimental disciplines and systematically reviewed all papers published at ICSE. Their findings revealed that statistical errors frequently occur in ESE, mirroring trends in medicine and psychology. Precisely, 30\% of the analyzed documents \textbf{lacked statistical hypotheses, did not calculate sample sizes, failed to check assumptions for statistical tests, and overlooked corrections for multiple tests}. These issues became more pronounced when experiments appeared solely in the validation section. The authors attributed these errors to “a lack of statistical training among researchers and an exploratory nature of much SE research.” They concluded that while other fields actively address statistical deficiencies, the software engineering community remains unaware of the problem.

Finally, in the same vein, recently, \cite{Vegas2024}, during ESEIW/ESEM 2024, dwelled on the fact that although SE has embraced empiricism, most of the studies are affected \textbf{by serious validity problems} on four critical dimensions: conclusion, internal, construct, and external validity.

More specifically, in \cite{vegas2023pitfalls}, they analyzed pitfalls in experiments with DNN in SE. The authors highlighted that statistical analyses often had problems such as unverified assumptions, inadequate reporting of descriptive and inferential statistics, and poor handling of multiple tests. For instance, 87\% of the studies reviewed \textbf{had incomplete attention to inferential statistics}, while 56\% \textbf{lacked necessary descriptive statistics}. Threats to internal validity were caused by a lack of control over extraneous variables and an inability to establish clear causal links. Besides, operationalization defects such as ill-defined definitions and misrepresentations of constructs make the treatments and response variables unreliable.

\cite{Vegas2024} warned the community of the possibility that specific methodological issues \textbf{would push SE to the situation faced by psychology, i.e., a reproducibility crisis}. She pointed out that steps toward improving empirical practices require improvements in experimental design, statistical rigor, and the principle of causality. Specific recommendations were strengthening methodological education, promoting transparency of data and methods, and building a culture of critical but constructive peer review to assure reliability in the research findings.

Our interest is driven by the critical need for in-depth analysis and the classification of good and flawed methodological pattern analysis in the whole field of SE. 
Our related work, albeit pioneering such an investigation, focused on specific communities, such as DNN for SE or ICSE. Thus, we are scratching the surface of a very deep iceberg that only a full dive can finally systematically address the daunting issues.

From the 2006 Dagstuhl seminar to the most recent findings in 2024, our community identified \textbf{methodological shortcomings}, such as statistical errors, inadequate reporting practices, generalisability, and consideration of validity threats. 

The findings from our pilot study and the workshop strongly confirm the urgent call for a serious rethink and reform of data analysis and reporting in SE. 

There is a need for a more rigorous and systematic approach to SE research's experimentation, analysis, and reporting strategies so that the results are reliable, reproducible, industrially relevant, and academically relevant. Our work is a step toward this foundation and toward establishing an SE research culture of methodological reflection and innovation.

%In light of our findings, we now aim to expand the scope of our investigation and the collected sample of papers. We will focus on a more in-depth analysis and categorization of the methodological pattern and the author's motivation behind good and flawed decisions. 

\section{Methodology Overview}
\label{sec:methods}
This section details the overall methodology of this empirical study, introducing the main goal and the adopted process. We design and conduct the empirical study according to the guidelines defined by Wöhlin et al. \citep{wholin2012experimentation}.

% \subsection{Goal}
The \textit{main goal} of our study is two-fold, as follows:  
\begin{enumerate} 
    \item[G$_1$] \textit{Investigate} the methodological statistical patterns for data analysis usage in ESE studies, distinguishing between adequate (``good'') patterns that adhere to best practices and inadequate (``bad'') patterns that compromise methodological rigor (as described in Section \ref{sec:Pattern}); 
    \item[G$_2$] \textit{Evaluate} the experts' inadequate methodological pattern detection and remediation capabilities. 
\end{enumerate}

Our \textit{perspective} is of researchers and practitioners in the ESE community. 

% \subsection{Study Context}
The study draws on a rich corpus of ESE literature spanning the last 30 years (1994–2023). These studies encompass various domains and methodologies, reflecting the evolution of empirical data analysis practices in ESE.

\begin{figure}
    \centering
    \includegraphics[width=0.75\linewidth]{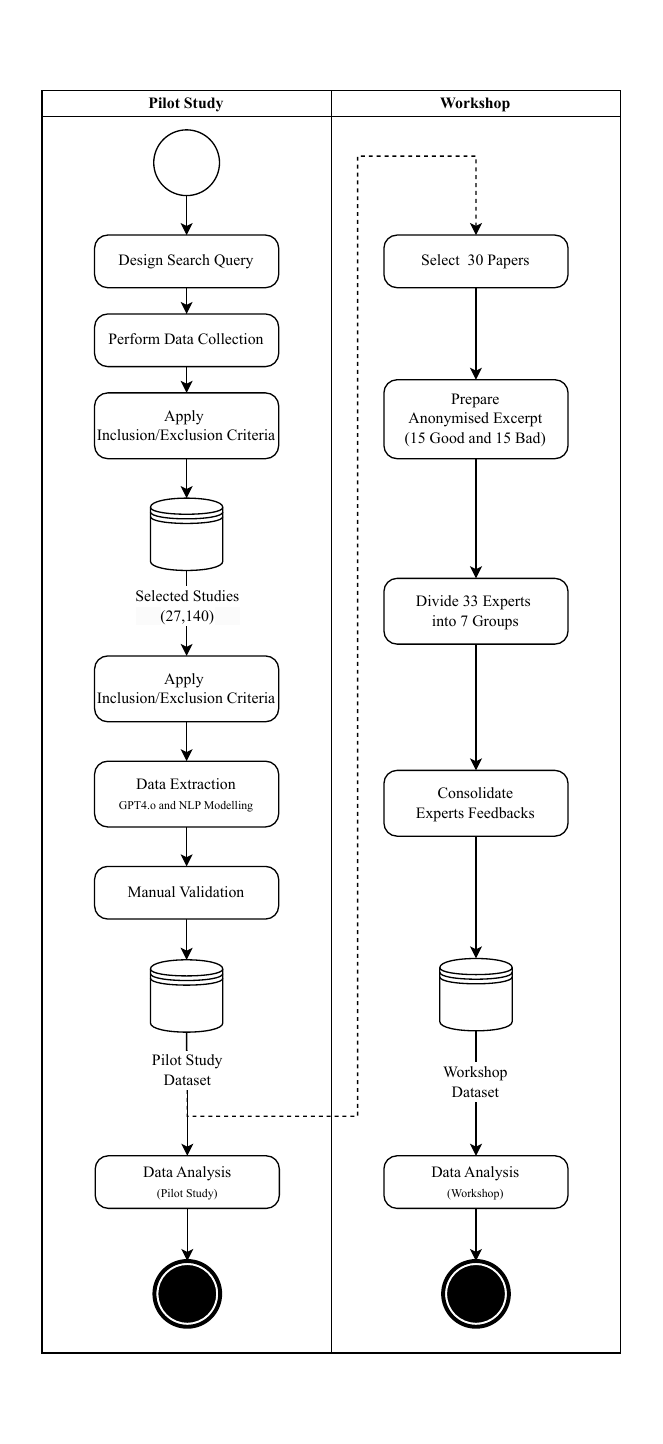}
    \caption{Study Setup and Data Collection Process}
    \label{fig:Process}
\end{figure}

% \subsection{Study Setup and Data Collection}
Figure \ref{fig:Process} shows the overall study setup and data collection process. We modeled our process via the Business Process Model and Notation (BPMN) 2.0. BPMN is a business process modeling standard \citep{bpmn2_specification}, which provides a graphical notation to intuitively describe a method to technical users and business stakeholders for clear communication and collaboration. 
To lay a solid foundation for the investigation, according to Figure \ref{fig:Process}, we split the study into two steps: 

\begin{enumerate}
    \item \textbf{Pilot Study} (Section \ref{sec:pilot}): Large-scale analysis of patterns and trends characterized by the methods of data analyses in the literature on ESE; 
    \item  \textbf{Workshop} (Section \ref{sec:workshop}): Involvement of ESE experts to verify and further inform findings from the pilot study.  
\end{enumerate}

\section{Pilot Study}
\label{sec:pilot}
This section presents the methodology, results, discussions, and threats to the validity of the pilot study.

\subsection{Study Design}
This section details the goal, research questions, data collection, and analysis. 

\subsubsection{Goal and Research Questions}
\label{subsec:goal1}

In the \textbf{pilot study}, we focus on the first goal (G$_1$), which is to investigate the use of methodological, statistical patterns for data analysis in ESE studies. We distinguish between adequate (``good'') patterns that adhere to best practices and inadequate (``bad'') patterns that compromise methodological rigor. 
Our \textit{perspective} is of researchers and practitioners in the ESE community.

Therefore, we define two Research Questions (RQs): 

\begin{boxC}{\textbf{$RQ_1$}}
  What is the historical trend of empirical data analysis methodology?
\end{boxC}

ESE relies heavily on rigorous methodologies to produce reliable, actionable results. However, the complexity of software engineering practices and the diversity of empirical approaches often lead to inconsistencies in study designs. These methodological issues can significantly affect the quality and reproducibility of research findings \citep{ccarka2022effort, esposito2024validate}. Stemming from our current peer-review experience and leveraging Dagstuhl seminar's legacy \citep{DBLP:conf/dagstuhl/2006esei} with this question, we aim to conduct a pilot study on state-of-the-art data analysis methodologies applied to past ESE studies. 

\begin{boxC}{\textbf{$RQ_2$}} 
  What are the most common methodological statistical patterns in empirical software engineering data analysis?
\end{boxC}
Empirical data analysis in software engineering often employs various methodologies, ranging from basic statistical tests to complex machine learning models. However, the diversity of approaches raises the question of whether there are recurring patterns or widely adopted practices that define the field. This RQ aims to identify these common methodology patterns to understand how researchers approach data-driven inquiries. This exploration will help highlight prevalent trends and potential inconsistencies or over-reliance on specific techniques, providing a basis for improving methodological guidelines and ensuring rigor in empirical studies.
Today's authors often reuse established beliefs from previous experts. As a result, if methodological issues remain challenging to detect, we risk perpetuating them like a virus rather than stopping them. Therefore, assessing the current expert capabilities is crucial to detect inadequate methodological patterns and suggest the correct remediation.

\subsubsection{Study Context}
This section describes the context in which our study is conducted. We first retrieved 504,062 articles from the Scopus database using a specific search query targeting a subset of the available ESE research based on their advertised empirical methodology (see Scope of the Search). After applying inclusion and exclusion criteria (detailed in Table \ref{tab:criteria}), we refined the dataset to 27,140. These studies formed the basis for identifying and characterizing methodological patterns via Natural Language Processing (NLP) and large language models (LLMs).

\subsubsection{Study Setup and Data Collection}
This section presents the data collection workflow for the pilot study.

\textbf{Overview}. Due to the vast corpus of over 27k papers, we aimed to extract the relevant information leveraging NLP and LLMs automatically. To identify the variable types, which is a trickier task than looking for text matching the name of statistical tests, we used ChatGPT-4 with a self-reflection prompting technique. The rest of the paper's text was normalized to remove diacritical symbols to allow a more straightforward NLP application to infer preprocessing methods, computational models, performance metrics, and post hoc tests. 

\textbf{Scope of the Search}. Our study is not a systematic literature review or mapping study. Thus, we deviate from the usual SLR guidelines \citep{DBLP:conf/icse/KitchenhamDJ04} while still safekeeping generalizability given the vast corpus of selected papers. Therefore, we chose only one search engine, Scopus, which allowed free access to its API to automate most of the literature search and papers gathering process. We employed the following search string:

\begin{boxCNoTitle}
\texttt{KEY (empirical) OR KEY (experiment) OR KEY (`case stud*'")}
\end{boxCNoTitle}
We limited the search to the SE domain and considered studies published between 1994 and 2023. We used the asterisk character (*) for the second term group to capture possible term variations, such as plurals and verb conjugations. We performed the data collection through official web APIs. We collected a total of 504,062 articles. We applied our inclusion and exclusion criteria (see Table \ref{tab:criteria}), excluding 476,922, thus focusing on the remaining 27,140 research articles from major computer science publishers such as IEEE, ACM, Elsevier, and Springer.%comprising 13,418 Elsevier, 5,298 Springer, and 8424 IEEE/ACM articles. 

\textbf{Text Extraction.}
We extracted text from the research papers' methodological sections to avoid confusion in analyzing sections unrelated to the paper's methodology. These sections were identified using key terms commonly associated with research methodology, such as “data analysis,” “method,” “design,” “result,” “discussion,” and “research question.”

\textbf{Determining Variable Type using LLM Prompting}. SLR requires multiple authors to read, extract, and synthesize information from a usually large set of papers \citep{esposito2024validate}. In our context, reading and manually coding over 27k papers was deemed unfeasible for a pilot study. Therefore, we decided to leverage LLMs for specific data extraction tasks. Users provide input through a question or a prompt to interact with such models. This prompt can be iteratively refined to enhance the model’s output and better align it with the user’s goals. In the context of LLMs, prompt engineering refers to the deliberate crafting and optimization of prompts to guide models in performing desired tasks effectively \citep{esposito2024large}. LLMs are pre-trained on extensive datasets, and prompting leverages this knowledge for task execution without requiring extensive additional training. Prompts translate tasks into structured inputs, enabling LLMs to generate relevant and accurate outputs. Prompting can be employed in zero-shot \citep{brown2020language}, few-shot \citep{xia2020low}, or fully supervised scenarios, e.g., conversational agents \citep{lee2023prompted,white2023prompt}, to achieve high performance across diverse NLP tasks. Specifically, we employed ChatGPT-4 to determine the variable type (e.g., text, nominal, binary, ordinal, interval, ratio, discrete, continuous, or time). We began with the following simple zero-shot prompt:
\begin{lstlisting}
{
    "role": "user" 
    "Content": "[extracted paper's section]. \n Question: Read the input text and determine the type of variable used for the data analysis in the research article?"
}
\end{lstlisting}

Applying this prompt to the same research article multiple times often led to inconsistent results for the inferred variable type. These inconsistencies revealed a key limitation of the basic zero-shot prompting approach: a mechanism to encourage the model to evaluate or reflect on its reasoning critically. To overcome this challenge, we implemented a self-reflection-based prompting technique \citep{wang2023plan}. This method explicitly guides the model to reflect on its reasoning process by:
\begin{itemize}
    \item Analyzing the input text thoroughly.
    \item Inferring the variable type while critically assessing its alignment with the listed types.
    \item Offering a factual explanation grounded solely in the content of the research article, avoiding any speculative reasoning.
\end{itemize}
The updated self-reflective prompt was structured as follows:

\begin{lstlisting}
{ "role":"user" 
    "Content" : "[extracted paper's section]. \n Question: What is the type of variable used for the data analysis in the research article? Infer the variable type, which must be one of text, nominal, binary, ordinal, interval, ratio, discrete, continuous, or time. Infer by self-reflecting on why/why not the variable type matches one of the listed types. Respond precisely based on the facts in the research article without hypothesizing. The response must be in a JSON format containing only 'variable_type'along with its inferred value and 'explanation'  containing a brief explanation of how the variable_type was inferred."
}
\end{lstlisting}

Reflection-based prompting introduces an additional layer of internal validation, prompting the model to critically evaluate its inferences before producing a response. This approach enhances robustness by prioritizing fact-based reasoning, ensuring that explanations are directly aligned with the input text, and reducing variability when the same content is analyzed multiple times. Leveraging this technique, we achieved consistent and factually grounded determinations of variable types across repeated evaluations.

To validate the model’s output, two authors independently reviewed a subset of paper classifications, achieving a 95\% confidence level with a 5\% margin of error.

\textbf{Text Normalization.} We perform text normalization once these sections relevant to data analysis are identified and extracted. First, we decompose the text containing diacritical accents into their simplest form, for instance, replacing ``\'e'' with ``e'' and the accent mark (´) with (\texttt{'}). This critical step cleans and standardizes the text to ensure consistency, preparing it for accurate keyword detection. Then, remove any characters that are not ASCII. 

\textbf{NLP Modeling.} The primary goal of NLP modeling is to infer the components of data analysis used in a research article. For each data analysis step, at a high level, we identify the type of preprocessing (if any), the name of the computational/statistical model used for analysis, its performance metric for evaluation, and finally, posthoc tests to determine where significant differences exist between groups after the initial analysis using the computational/statistical model. To accomplish the NLP modeling,  we rely on our taxonomy (see Section \ref{subsec:dat}) containing a dictionary of keywords to categorize common preprocessing techniques, computational models, performance metrics, and post-hoc tests used in the research article. We utilize the counts of occurrences of these terms in the normalized text to ascertain which technique, model, metric, or post-hoc test is employed for each category. If there are different entries for each category, then the different entries for each category are included in our database. Each research article's unique identifier, category of analysis, and detected keywords are recorded. Even if no relevant keywords are found, this outcome is logged to ensure every article is accounted for. Once all research articles are processed, i.e., evaluated for the frequency analysis of all the categories from the taxonomy, we begin our analysis.

\begin{table}
\centering
\caption{Inclusion and Exclusion Criteria}
\label{tab:criteria}
\begin{tabular}{p{0.1cm}p{10cm}}
\hline
\multicolumn{2}{c}{\textbf{Inclusion Criteria}}                                                                                                                          \\ \hline
I1. & The study relates to the field of Software Engineering.                                                                            \\
I2. & The study presents a clearly defined methodological section.                                                                                                             \\
%I3. & The study is a full paper longer than six pages.                                                                                                                   \\
%I4. & The study provides a replicability dataset that is either reachable or provided after asking to authors.                                                           \\ 
\hline
\multicolumn{2}{c}{\textbf{Exclusion Criteria}}                                                                                                                          \\ \hline
%E1. & Solely a literature review or survey article.                                                                                                                      \\
E1. & Non-peer-reviewed academic literature.                                                                                                                             \\
E2. & Book chapters or dissertations.                                                                 \\
E3. & Studies not written in English.                                                                                                                                    \\
E4. & Studies whose full text is unavailable.                                                                                                                            \\
E5. & Studies published to a venue unrelated to the discipline of Computer Science.                                                                                      \\ \hline 
%E7. & Studies published to a journal or conference with a CORE ranking of less than A and H-index less than 40, and that have a citation count of less than 20. \\ \hline
\end{tabular}%

\end{table}

\subsubsection{Data Analysis}
This section presents our data analysis procedure to answer our RQs.
To answer $RQ_1$, we queried our database and hierarchically grouped the data. Therefore, we analyze the historical trends in empirical data analysis (EDA) methodology. To answer $RQ_2$, i.e., the common methodology patterns in EDA, we leveraged Sankey Plots (SP) \citep{doi:10.1680/imotp.1898.19100}. To analyze 30 years of data analysis methodologies and issues, SP can be a powerful tool for visualizing the evolution and distribution of methods, highlighting flows and transitions over time \citep{doi:10.1680/imotp.1898.19100}. SP reveals patterns, significant shifts, and trends in data analysis practices by mapping the movement of methods from their inception to adoption or decline across decades. More specifically, we can show how paper falling into a specific category of the taxonomy is later distributed to the subsequent categories, like water in a pipe system.

\label{sec:pilot:methodology}

\subsection{Results}
This section presents the early findings from our pilot and the workshop. 

\subsubsection{Historical Trends in Empirical Data Analysis Methodology (\texorpdfstring{RQ$_1$}{RQ1})}
\begin{figure}
    \centering
    \includegraphics[width=0.75\linewidth]{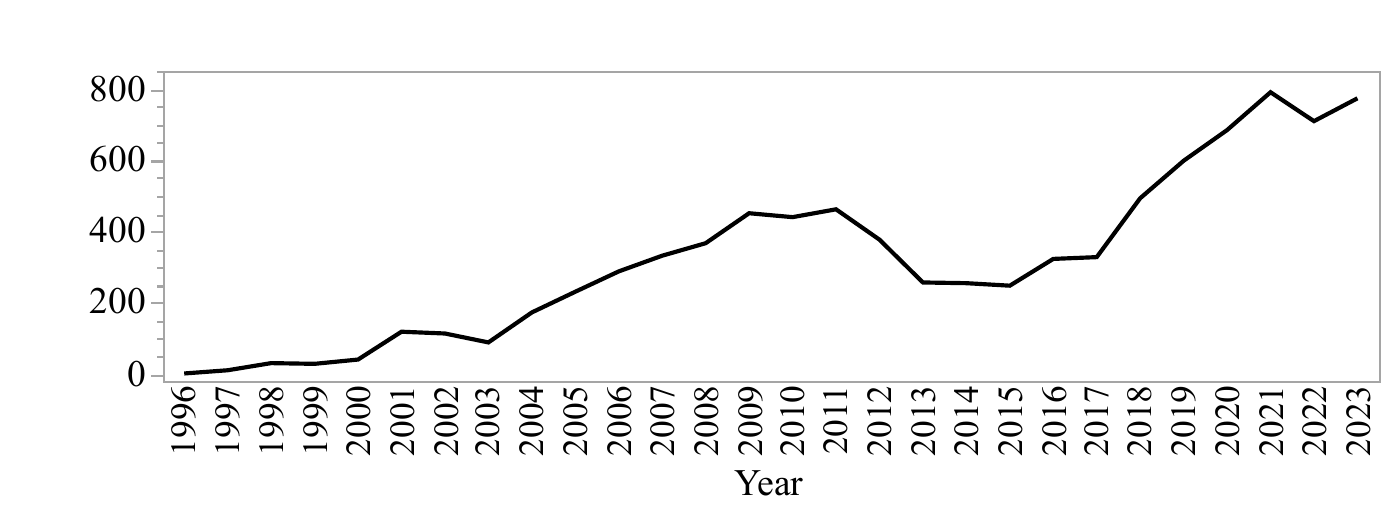}
    \caption{Empirical Software Engineering Publication Trend}
    \label{fig:pubtrend}
\end{figure}

Figure \ref{fig:pubtrend} shows the ESE publication trend in the selected venues. According to Figure \ref{fig:pubtrend}, interest in our field has increased steadily since the mid-90s, reaching more than 800 papers published in a year (2023). Following this preliminary trend, we are interested in analyzing in depth the trend of the methodological choices in terms of Variable type, data distribution checks, and hypothesis testing.

\begin{figure}
    \centering
    \includegraphics[width=0.75\linewidth]{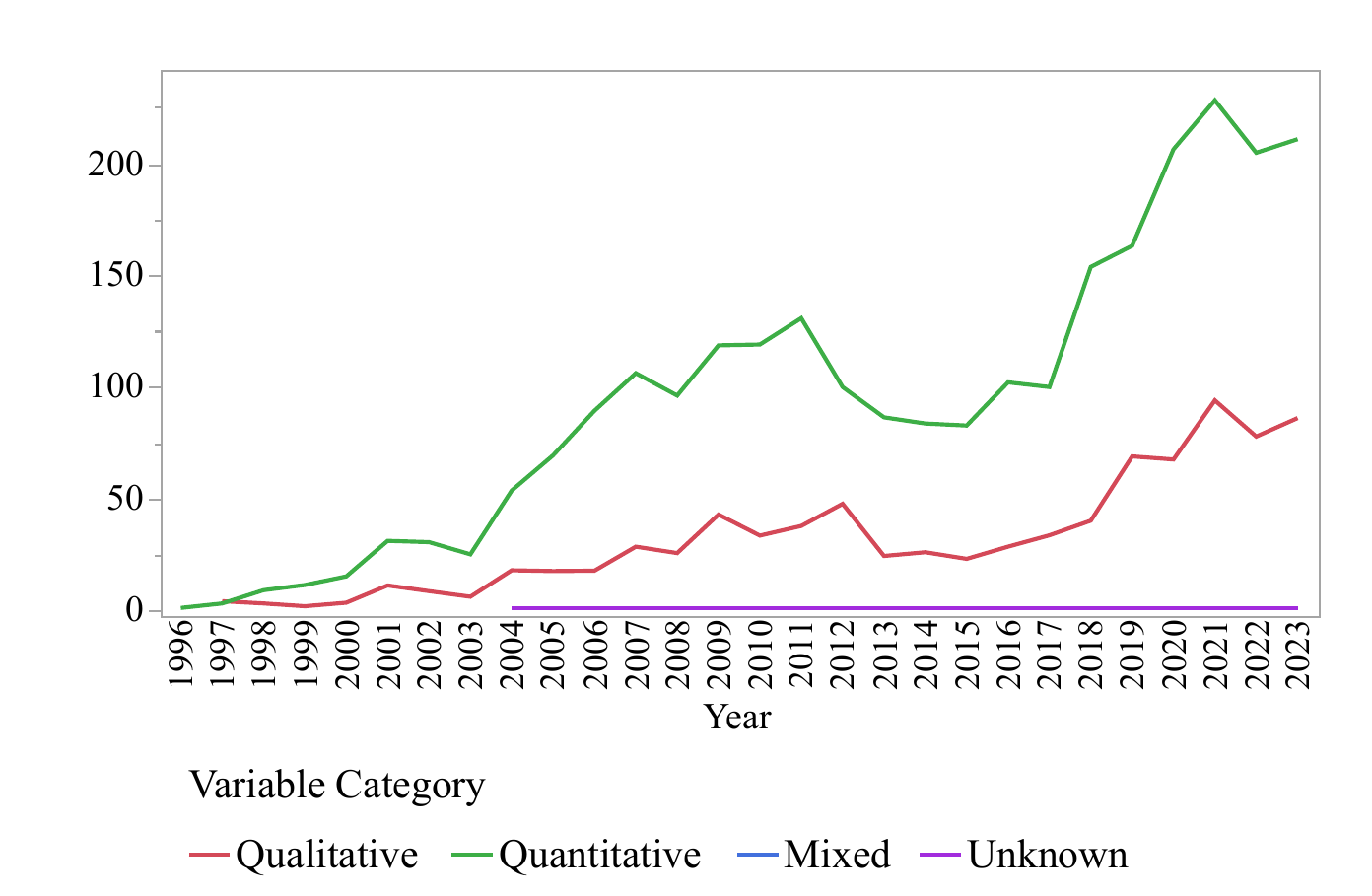}
    \caption{Variable Type Trend}
    \label{fig:vartypetrend}
\end{figure}

Figure \ref{fig:vartypetrend} presents the study's variable type trend. According to Figure \ref{fig:vartypetrend}, quantitative studies are consistently the most preferred analyzed variable with a consistent gap from qualitative ones. Conversely, the variable under examination was negligible in mixed and unclear studies.

\begin{figure}
    \centering
    \includegraphics[width=0.75\linewidth]{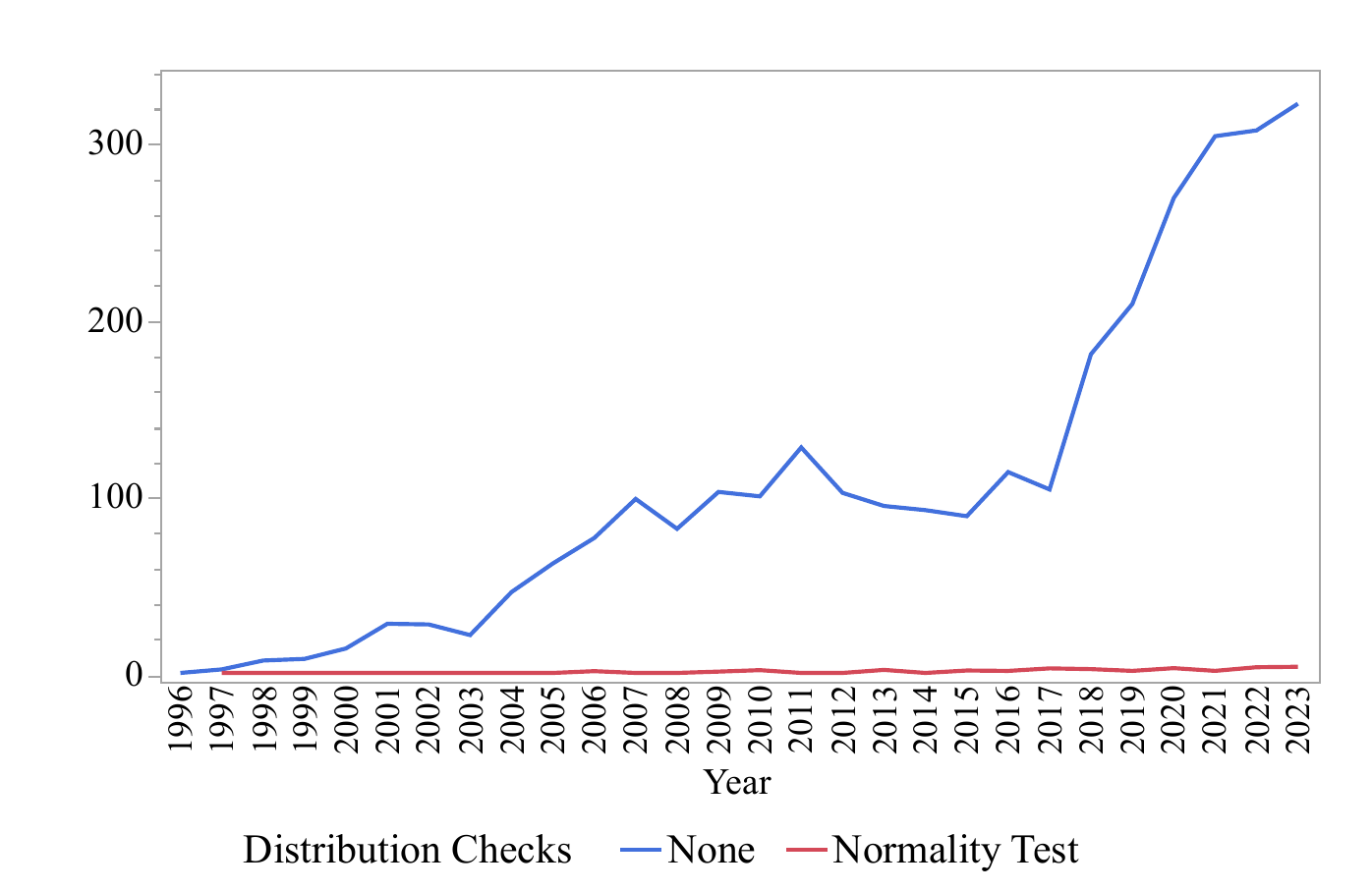}
    \caption{Distribution Check Trend}
    \label{fig:distrtrend}
\end{figure}

Furthermore, Figure \ref{fig:distrtrend} presents the study's variable distributional checks trend. According to Figure \ref{fig:distrtrend}, authors rarely check for the distribution of the variables, and among the few that do that, a very handful of them check for normality. Normal distribution is a prerequisite for specific hypothesis testing and analysis methods. Not checking the distribution and applying an erroneous test threaten the entire validity of the study.

\begin{figure}
    \centering
    \includegraphics[width=0.75\linewidth]{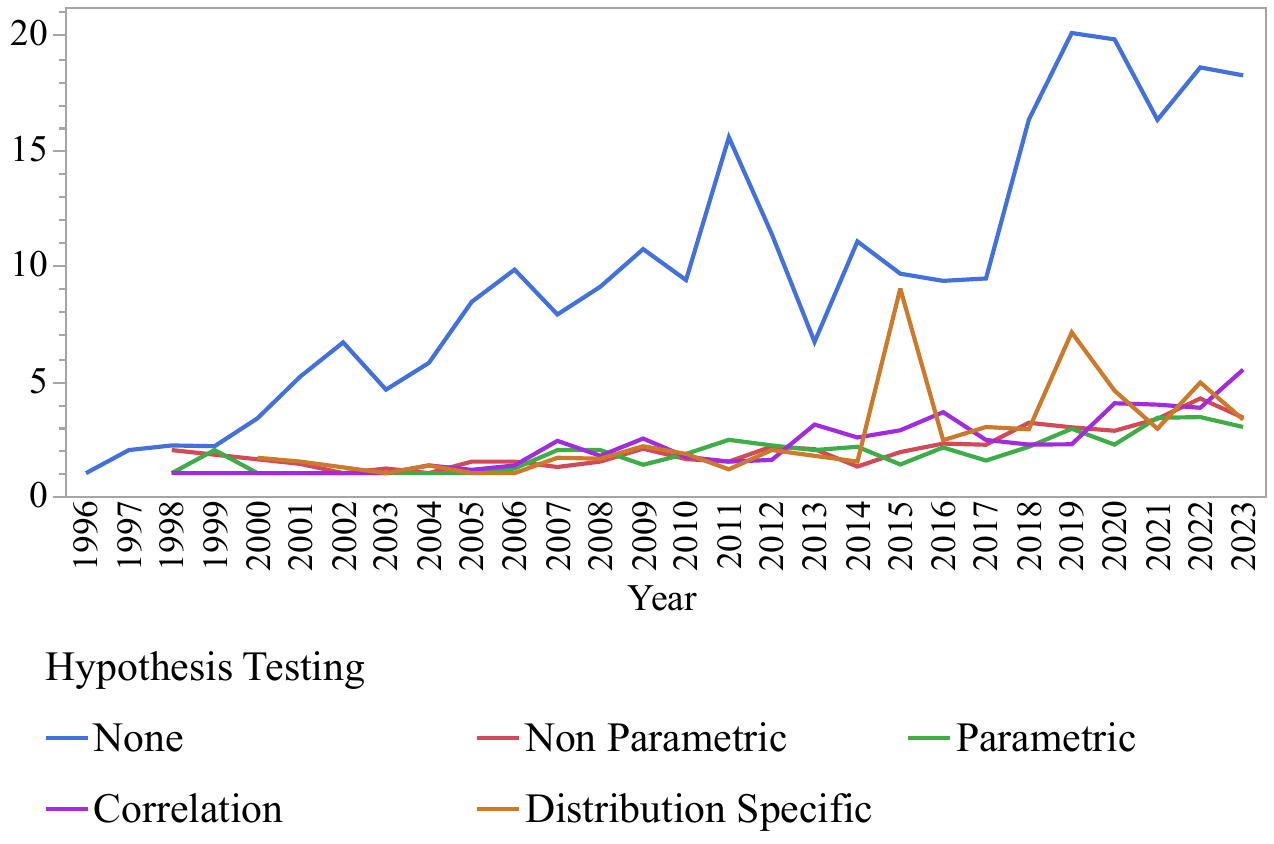}
    \caption{Hypothesis Testing Trend}
    \label{fig:hypotesttrend}
\end{figure}

Finally, Figure \ref{fig:hypotesttrend} presents the study's hypothesis-testing trend. According to Figure \ref{fig:hypotesttrend}, most works do not test the hypothesis. Most studies that test their hypothesis use correlations or non-parametric or distribution-specific tests.

Therefore, we can observe that over the last decade \textbf{authors have focused on quantitative study, rarely checking the distribution of the collected data or testing their hypothesis}.

\subsubsection{Common Methodology Patterns in Empirical Data Analysis (\texorpdfstring{RQ$_2$}{RQ2})}

\begin{figure}
    \centering
    \includegraphics[width=1\linewidth]{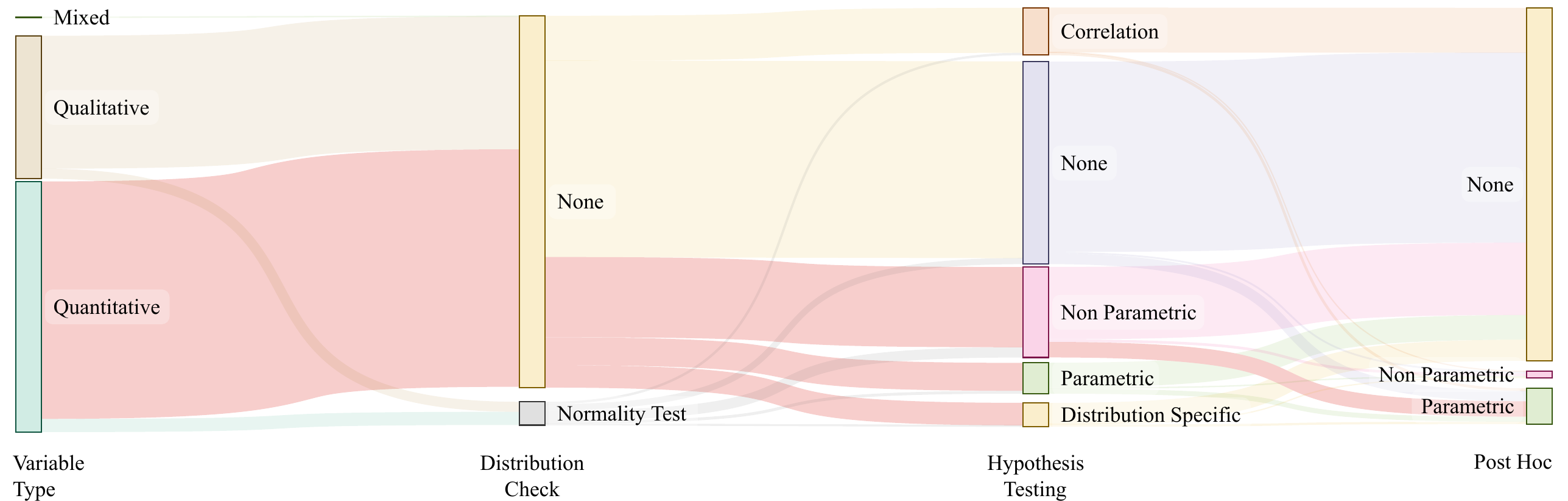}
    \caption{Examples of inadequate methodological statistical patterns, highlighted in pale pink.}
    \label{fig:sanky}
\end{figure}

Figure \ref{fig:sanky} presents the Sanky Diagram, highlighting three examples of inadequate patterns. According to Figure \ref{fig:sanky}, roughly 90\% of the papers focusing on quantitative or qualitative data did not check the distribution of the collected data. We refer to this set of papers as \textbf{$\mathcal{A}$}. About 10\% of $\mathcal{A}$ proceeded with parametric tests, which require data to be normally distributed, and 90\% of them did not employ any post hoc correction.

Similarly, About 10\% of $\mathcal{A}$ proceeded with non-parametric tests, which do not require data to be normally distributed, and then circa 15\% of these last ones use a parametric post hoc correction, which, in turn, required data normally distributed.

Finally, 15\% of \textbf{$\mathcal{A}$} performed a distribution-specific test without being aware of the data distribution.

Therefore, we can observe that \textbf{different methodological issues affect the last 30 years of ESE}.

\label{sec:pilot:results}

\subsection{Discussions}
This section discusses our findings. This study is a pilot for our broader, ongoing evaluation of state-of-the-art literature across all digital libraries, extending beyond the current paper selection and the investigation scope.

The analysis of the publication trends in ESE, i.e., RQ$_2$ gives several interesting insights about the development of methodological practices within the field. The interest of scholars has increased steadily over the years, culminating in a record number of publications in 2023, indicating a living and growing research community. However, taking a closer look at the methodological trends reveals huge gaps. Quantitative studies dominate the landscape, favoring measurable and numeric data, while qualitative studies are notably underrepresented. This may indicate a preference in the subject area for precision and generalizability over exploratory or contextual insights.

The trend realized in the checking of the distribution of data is worrying. Very few checked the distribution of their data; out of them, an even smaller fraction checked normality. Given that many statistical tests are based on the assumption of normal distribution, this step, if ignored, may lead to the failure of the results. Similarly, the trend in hypothesis testing shows that although some studies do this important thing, many do not, and most lean toward simpler analyses of correlation or non-parametric tests. This indicates a limited engagement with robust statistical methods that may reduce the robustness and reliability of conclusions made.

These trends indicate the need for more rigorous methodological thresholds in ESE research. Such checks on distribution should be comprehensive; encouraging a culture of robust hypothesis testing will increase the credibility and impact of future studies.

More specifically, RQ$_2$ reveals that over the last 30 years, various methodological issues have impacted the ESE field. In particular, problems such as the blind application of tests without considering their prerequisites, i.e., testing the distribution and shape of the data, threaten the validity and generalizability of many ESE studies.
\label{sec:pilot:discussions}

\subsection{Threats to Validity}
This section discusses the threats to the validity of our pilot study, categorized into construct, internal, external, and conclusion validity, following established guidelines \citep{wholin2012experimentation}.

\textbf{Construct validity} focuses on whether our measurements accurately reflect what we claim to measure \citep{DBLP:journals/ese/RunesonH09, Basili1994}. One significant threat arises from the use of a single search engine, Scopus, and a specific search string that included terms like “empirical,” “experiment,” and “case stud*.” While this search string aimed to be broad enough to capture a wide range of ESE studies, it likely excluded relevant research that did not explicitly use these keywords excluding other empirical methodologies such as the ones presented in Section \ref{sec:BG}. This limitation could reduce the comprehensiveness of our dataset. Additionally, the predefined taxonomy used in our NLP modeling was developed with input from domain experts. However, this taxonomy may not fully capture the diverse methodologies employed in empirical software engineering studies, potentially leading to misclassification or oversight. Moreover, threats to construct validity may also arise from the behavior of the researchers conducting the study or the NLP tools themselves, potentially biasing the results \citep{wholin2012experimentation}. We will address those threats in the upcoming in-depth literature review on all the empirical study methodologies and the refinement of our taxonomy. 

\textbf{Internal validity} concerns how our study design identifies causal relationships between variables \citep{wholin2012experimentation}. Our reliance on ChatGPT-4 for determining variable types introduces a notable risk. Although we applied a self-reflection prompting technique to improve reliability, the inherent limitations of large language models may still result in errors or inconsistencies in the output. Similarly, the text normalization steps, such as removing diacritical marks, may have inadvertently altered or obscured key methodological details. While necessary to streamline the analysis, these preprocessing choices could compromise the results' accuracy. Potential instrumentation issues, such as errors in the tools or techniques used to extract and analyze the data, also represent a source of bias \citep{wholin2012experimentation}.

\textbf{External validity} examines how well our findings generalize beyond the specific dataset used in the pilot study \citep{wholin2012experimentation}. While we analyzed over 27,000 articles, the decision to focus solely on Scopus-indexed studies published between 1994 and 2023 may have excluded important works from other databases, non-English publications, or gray literature. Consequently, our findings may not fully represent the broader landscape of ESE research. Moreover, although the dataset spans a wide range of methodological fields, some less common approaches may have been underrepresented, which could skew the observed trends and patterns. The small number of experts consulted during the taxonomy design also limits the generalizability of our results, although we sought to include a diverse set of empirical fields.

\textbf{Conclusion validity} addresses the reliability of the inferences drawn from the data \citep{wholin2012experimentation}. Since this study was a pilot, likely, not all methodological patterns were fully captured. This limitation may lead to premature conclusions about historical trends or prevalent practices in empirical data analysis. Furthermore, while the second and third authors independently reviewed a subset of classifications to validate the LLM outputs, the large corpus size prevented manual validation of the entire dataset. This reliance on automated processes could affect the accuracy and robustness of our conclusions. Our conclusions also rely heavily on the specific metrics chosen for accuracy, and alternative metrics might reveal additional insights \citep{esposito2024leveraging, esposito2024validate}.

%Additional threats to validity include the potential for human bias in interpreting the results of NLP and LLM analyses. In cases where model outputs required manual adjustments or clarifications, subjective judgments may have influenced the final interpretation. Moreover, given the pilot nature of the study, certain methodologies, such as the self-reflection prompting for LLMs, were applied without extensive iterative refinement. These factors could affect the overall reliability and depth of the insights drawn from the study. By acknowledging these threats and their potential impact, we aim to provide a transparent and grounded discussion of the study's limitations while laying the groundwork for future refinement and expansion.
\label{sec:pilot:threats}

\section{ESE Experts' Evaluation}
\label{sec:workshop}
This section presents the methodology, results, discussions, and threats to the validity of the workshop.

\subsection{Study Design}
This sub-section details the goal, research questions, data collection, and analysis.

\subsubsection{Goal and Research Questions}
\label{subsec:goal2}
In the \textbf{workshop}, we focus on the second goal (G$_2$)  in which we \textit{evaluate} the experts' inadequate methodological pattern detection and remediation capabilities. Our \textit{perspective} is of researchers and practitioners in the ESE community. Therefore, we defined the last two RQs: 

%We investigate and evaluate whether ESE specialists recognize lawed methodological patterns suggesting the right approach to use. 

\begin{boxC}{\textbf{$RQ_3$} }
How accurate are the ESE experts at detecting inadequate methodological statistical patterns for data analysis?
\end{boxC}
Despite the critical role of accurate statistical analysis in empirical research, the potential for misuse remains a significant challenge. $RQ_3$ focuses on evaluating how well ESE experts can identify instances of statistical misuse in empirical data analysis.  Finally, we must determine whether experts can suggest the correct remediation for the inadequate methodological statistical patterns for data analysis. Hence, we ask:

\begin{boxC}{\textbf{$RQ_4$}}
How accurate are the ESE experts in suggesting remediations to inadequate methodological statistical patterns for data analysis?
\end{boxC}
Identifying statistical misuse is only part of the solution; proposing effective remediation is equally crucial for upholding the integrity of research findings. $RQ_4$ examines the ability of ESE experts to suggest accurate and practical solutions for correcting statistical issues in empirical analyses. This question assesses whether experts possess the diagnostic skills to identify problems and the prescriptive knowledge to guide improvements.

\subsubsection{Study Context and Population}
This section describes the context in which our study is conducted and the characteristics of the workshop participants.

\textbf{Study Context}
Building on the insights from the pilot, we then conducted a workshop with \expertNo experts in ESE to test their capability to identify and correct methodological issues. Experts were selected based on their extensive experience in empirical research and review of methods within the software engineering domain. The data from the participants were collected in a way that is considered ethical and in line with the GDPR act on data collection and protection. The study also followed the ACM policy on human participant research \citep{ACM_Human_Participants_2021}.

\textbf{Participant Demographics and Expertise}. We identify three points, as follows: 
\begin{itemize}
    \item \textbf{Experience}: Participants included senior researchers, industry practitioners, and peer reviewers with 10–25 years of experience in ESE.
    \item \textbf{Fields of Expertise}: The participants specialized in experimental design, statistical analysis, and methodological rigor within software engineering.
    \item \textbf{Global Representation}: Experts from diverse geographical regions were selected to reflect varied academic and industrial perspectives.
\end{itemize}

\textbf{Ethical Considerations}
All participants were briefed on the study's objectives and provided informed consent before participation. The workshop involved anonymized research excerpts to mitigate bias, ensuring that neither author names nor publication details influenced the assessments.

\subsubsection{Study Setup and Data Collection}
According to Figure \ref{fig:Process}, regarding the workshop, we have selected 30 papers (that contain 15 adequate methodological patterns and 15 inadequate methodological patterns, as reported in Section~\ref{sec:Pattern}) from the collected ones in the pilot and extracted only the methodology and data analysis sections with no reference to the authors and the actual paper to avoid bias. We prepared a task overview and gathered \expertNo experts split into seven groups. To each group, we provided four excerpts, two of which had inadequate practices and two of which had adequate practices. We asked the experts to read the excerpts and highlight any methodological issues they may find according to the provided taxonomy.

\subsubsection{Data Analysis}
This section presents our data analysis procedure to answer our RQs. To create the Ground Truth (GT) for our study, we classified the 30 selected paper excerpts using our taxonomy (see Section \ref{subsec:dat}) to identify adequate and inadequate practices. We provided the experts with a copy of our taxonomy categories. Along with the taxonomy, we shared the selected paper excerpts and asked the experts to assess the practices described in these excerpts according to our taxonomy. Specifically, we requested that they evaluate each practice’s adequacy or inadequacy based on the categories defined in our taxonomy. We needed to ensure their assessments aligned with the structure and criteria we used to create the GT.

To answer $RQ_3$, we collected and manually coded the expert's answers in our workshop and compared them against the GT.  To ensure consistency, the first two authors independently reviewed and refined the ESE expert's classifications through iterative coding, prioritizing more general keywords from the taxonomy when needed. In case of a disagreement, we discussed it and reached a consensus with the last author. To measure the expert's performance, we selected the Accuracy, Precision, Recall, F1-Score, and Matthews Correlation Coefficient (\textbf{MCC}), metrics that are commonly used in Information Retrieval (IR) tasks \citep{falessi2023enhancing} and have been successfully used in assessing human experts \citep{esposito2024leveraging}.

Finally, to answer $RQ_4$, from the collected answers, we classified as ``good suggestions'' [``bad suggestions''] an expert answer that refers to a specific keyword, model, or statistical test as correctly employed by the study, and the GT-backed up [did not back up] the expert's claim  \textbf{or} the expert suggested a good practice that was indeed good [was not a good practice]. Moreover, we also added the keyword ``Other Suggestions'') and classified them accordingly for each category in the taxonomy for expert suggestions. We then compared the overall assessment of the paper against the ground truth to extract accuracy metrics.

\label{sec:workshop:methodology}

\subsection{Results}
This section presents the early findings from our pilot and the workshop. 
\subsubsection{Accuracy of ESE Experts in Detecting Inadequate Methodological Patterns (\texorpdfstring{RQ$_3$}{RQ3})}
\begin{figure}
    \centering
    \includegraphics[width=\linewidth]{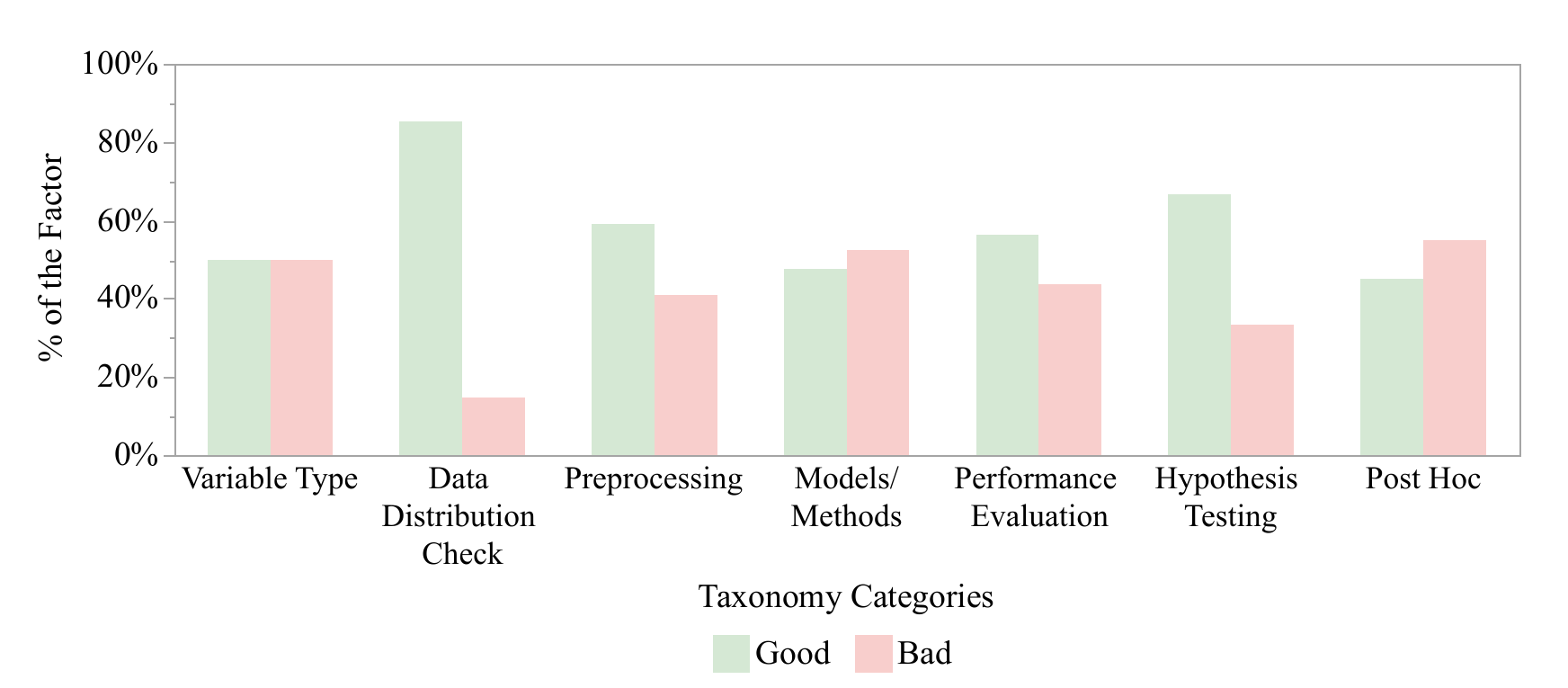}
    \caption{Adequate and Inadequate Methodological Statistical Patterns Distribution Suggestions from Experts by Category}
    \label{fig:summary}
\end{figure}

Figure \ref{fig:summary}  presents the distribution of adequate and inadequate suggestions from the experts summarized by the taxonomy category. According to Figure \ref{fig:summary}, the experts did not reliably spot methodological flaws. On the one hand, the experts could not reliably identify the variable types, appropriate models or methods, and post hoc tests needed for the study. This led to misclassification, i.e., labeling poor analyses as good and vice versa. On the other hand, the experts reliably classified the data distribution checks correctly. 
Therefore, we can observe that, on average, \textbf{ESE experts show limited precision in accurately reviewing methodologies based on our taxonomy}.

\subsubsection{ESE Experts’ Accuracy in Recommending Remediations for Inadequate Methodological Patterns (\texorpdfstring{RQ$_4$}{RQ4})}

\begin{figure}
    \centering
    \includegraphics[width=\linewidth]{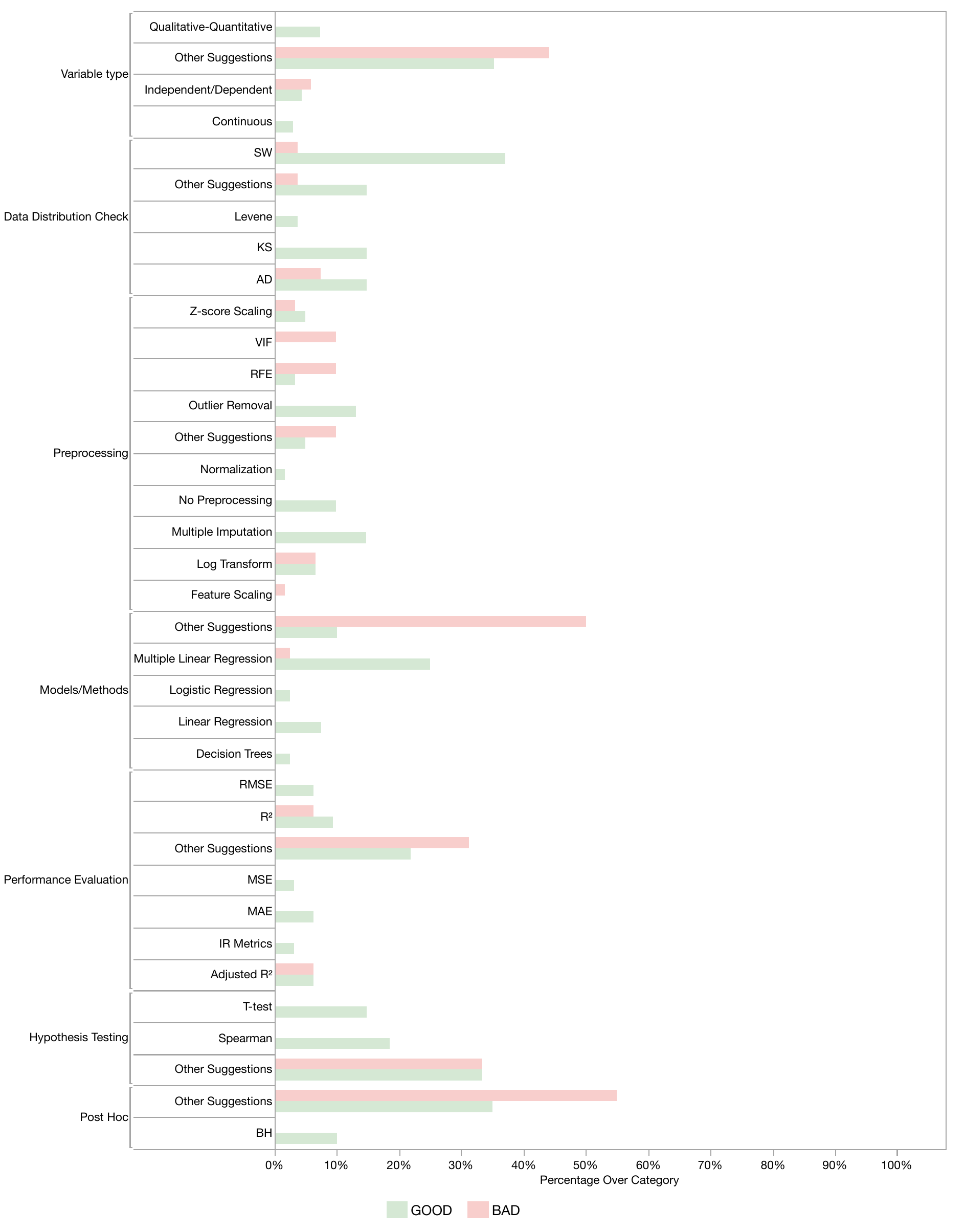}
    \caption{Distribution of Adequate and Inadequate Suggestions from Experts by Answer}
    \label{fig:synth}
\end{figure}

Figure \ref{fig:synth} shows the distribution of inadequate and good guesses from the experts by specific answers. According to Figure \ref{fig:synth}, referring to the findings in Figure \ref{fig:summary}, most of the errors the expert committed during the evaluation are in the other suggestions keyword. Hence, in most cases, when the expert suggested a different technique, model, or statistical test, the suggestion itself was wrong.

Finally, according to our results on the accuracy metrics computed for the expert across the seven groups and the provided scenarios with the overall evaluation of the papers provided, we note that the experts underperformed in each accuracy metric. 

More specifically, experts scored very low accuracy in terms of Recall (0.3846), Precision (0.3846),  Accuracy (0.3600),  F1-Score (0.3846), and a negative MCC (-0.2821).

Therefore, we can observe that, \textbf{on average, ESE experts could not identify statistical methodological issues}.

\label{sec:workshop:results}

\subsection{Discussions}
This section discusses our findings. This study is a pilot for our broader, ongoing evaluation of state-of-the-art literature across all digital libraries, extending beyond the current paper selection and the investigation scope.
According to our findings in RQ$_3$ and RQ$_4$, we showed that the \textbf{ESE experts confounded statistics principles} in reviewing the excerpts from the paper.  According to Figure \ref{fig:synth}, experts often wrongly suggested new statistical tests or different models and techniques. Moreover, our findings further highlight significant flaws in the peer-review process, as experts frequently misidentify appropriate statistical tests, models, and methods, thus confirming the previous findings of \cite{reyes2018statistical, Vegas2024}. Therefore, considering all these findings and the fact that our study stemmed from our own peer-review experience and the legacy of the 2006 Dagstuhl seminar, we can affirm that the problem exists and is unfortunately widespread all across our community.

\label{sec:workshop:discussions}

\subsection{Threats to Validity}
In this section, we discuss the threats to the validity of our observational study. We categorized the threats in Construct,  Internal, External, and Conclusion validity following established guidelines \citep{wholin2012experimentation}.

\textbf{Construct Validity} concerns how our measurements reflect what we claim to measure \citep{wholin2012experimentation}. Our design choices, measurement process, and data filtering may impact our results. To address this threat, we based our selection on past studies and well-established guidelines in designing our methodology \citep{DBLP:journals/ese/RunesonH09,Basili1994}. Moreover, threats to the construct validity may arise due to the behavior of both participants and researchers conducting the study. The participants may act differently merely because they know they are being studied \citep{wholin2012experimentation}. 

\textbf{Internal Validity} is the extent to which an experimental design accurately identifies a cause-and-effect relationship between variables \citep{wholin2012experimentation}. Regarding the focus group, we focused on the following two threats: (i) \textit{Instrumentation} refers to the potential impact of the tools or materials used during the study, in this case, the material for the focus group. A poorly designed task can compromise the reliability of the study’s results \citep{wholin2012experimentation}. To address this threat, we developed the task for the focus group with only direct questions, minimizing the need for interpretation and reducing the risk of misunderstandings that could lead to irrelevant answers. (ii) \textit{Maturation} refers to the possibility that participants’ responses may change over time \citep{wholin2012experimentation}. In our case, the task was completed in 15 minutes in person.

\textbf{External Validity} concerns how the research elements (subjects, artifacts) represent actual elements \citep{wholin2012experimentation}. Our sample size, albeit small, includes renowned experts in the ESE community spanning the most prominent methodological fields. As such, we tried to mitigate this threat by diversifying our interviewed experts' specific empirical field and methodology expertise.

\textbf{Conclusion Validity} focuses on how we draw conclusions based on the design of the case study, methodology, and observed results \citep{wholin2012experimentation}. 
Our conclusions rely on the specific accuracy metrics chosen, and there may be other aspects or dimensions of performance that we did not consider. To address this potential limitation, we selected Accuracy, Precision, Recall, F1-Score, and MCC to evaluate the human experts \citep{esposito2024leveraging,esposito2024validate}. 
\label{sec:workshop:threats}

\section{Insights and Limitations}
\label{sec:insightsAndLimitations}
This section presents the major limitations and an overview of the lessons learned from the overall study. 

\subsection{Lessons Learned}

On the one hand, the pilot study highlighted and reinforced past studies' results, thus motivating a more in-depth systematic study. On the other hand, the experts' underperformance shed light on an evident negative loop, i.e., the study's methodological quality is decreasing alongside the experts' preparation and accuracy.

\begin{keyTakeAways}[Insufficient Data Checks]
To ensure that research at ESE holds, more consideration ought to be given to ensuring, through stringent checks, that data is appropriate or valid.
\end{keyTakeAways}

Though the number of publications has increased over time, the major oversight on methodology is that few researchers have tested their data distribution. Such laxity may result in the wrong application of the parametric test, which assumes the normalcy of the data distribution. Gaps of this nature can invalidate research findings and question their validity. Embedding proper data distribution checks into standard routines will enhance the reliability and robustness of any study, furthering scientific rigor within the field.

\begin{keyTakeAways}[Bias in Methods]
Most quantitative studies depict a particular bias, reducing methodological variance within ESE.
\end{keyTakeAways}
While providing precision and generalization, quantitative studies override the rich insights that can be achieved from qualitative or mixed-method approaches. This trend illustrates a preference toward numerical analysis rather than contextual or exploratory understanding and thus might reduce the depth of research insight. Such methodological diversity-enabling qualitative research and mixed methods promote comprehensive insights about complex software engineering phenomena and will enrich the total body of knowledge.

\begin{keyTakeAways}[Lack of Hypothesis Testing]
The absence of hypothesis testing reduces the depth and reliability of conclusions in ESE studies.
\end{keyTakeAways}
The results show that hypothesis testing, considered one of the most essential elements in empirical research, is either missing or inadequately conducted in most studies. Most researchers cannot perform simple correlation analyses or non-parametric tests without post hoc corrections. This procedure narrows the interpretive possibilities of the research and calls into question its statistical robustness. Strong hypothesis testing and appropriate use of statistical methods will promote the validity of studies and a culture of critical scientific inquiry within the ESE community.

\begin{keyTakeAways}[Methodological Training Gaps]
The methodological flaws of the last three decades have created an urgent need to raise the level of research training.
\end{keyTakeAways}
The pilot study highlighted several critical issues, such as a general misuse of parametric tests and post hoc corrections that were largely unconsidered. Moreover, in the expert evaluations, Recall, Precision, and F1-Score were alarmingly low, whereas the negative values of MCC underlined the gross misunderstanding of basic statistical principles. These findings reveal an urgent need for targeted educational programs on statistical methodologies, emphasizing the correct application within empirical research. To date, this lacuna has serious repercussions on improving the reliability of expert reviews, consequently impacting the quality of ESE studies.

Workshops, tutorials, and detailed guidelines concerning methodological issues will provide researchers with the necessary opportunity to acquire skills in the proper design and execution of studies. Strengthening methodological rigor is imperative to ensure that the quality of the field improves to a level beyond which the long-term credibility of the research in ESE is ascertained.

\begin{keyTakeAways}[Expert Review Inaccuracies]
Misclassification of statistical methodologies by ESE experts has shown significant gaps in the accuracy of expert reviews.
\end{keyTakeAways}
These results show that the experts regularly failed to identify variable types, models, and statistical methods correctly. Misclassifications yield poor evaluations: poor analyses are classified as good and vice versa. Although experts fared better in identifying checks for data distribution, overall accuracy remains disconcertingly low. This raises a critical problem in the integrity of peer-reviewed research and a dire need for more comprehensive frameworks and training to enhance expert precision in evaluating methodologies.

\begin{keyTakeAways}[Inadequate Remediation Approaches]
Experts struggle to suggest the appropriate remediation strategies for the flawed methodological patterns and sometimes aggravate the problem.
\end{keyTakeAways}
Experts frequently recommend inadequate alternatives when asked to suggest alternative techniques, models, or statistical tests. These bad suggestions worsen the research's methodological flaws, possibly leading to a lack of valid conclusions. This lesson underlines the importance of clear-cut and standardized guidelines and supporting instruments that enable the expert to make proper and correct remediation choices.

\begin{keyTakeAways}[Peer Review Weaknesses]
The systemic flaws in experts' assessments may hinder the peer-review process.
\end{keyTakeAways}
The expert's underperformance may introduce and threaten the peer-review mechanisms. Experts could not spot critical statistical errors, such as improper hypothesis testing or inappropriate use of models. This indicates a systemic problem affecting the peer-review process since reviewers may not have the appropriate tools or the knowledge to conduct good reviews. Overcoming this flaw will involve embedding methodological checklists, training programs, and automated tools to help reviewers identify and address such issues.

\begin{keyTakeAways}[Community Reform Needed]
Widespread methodological challenges require community-driven reforms in ESE research practices and review standards.
\end{keyTakeAways}
These methodological flaws found in our study are not isolated but part of the general problem of the ESE community. As Prof. Sira Vegas pointed out, these problems persist across different research areas and peer-review processes. This challenge calls for collaboration to establish community-wide standards through workshops, methodological training, and best practices that raise the bar on high-quality research and accurate reviews.

\subsection{Limitations}

\begin{keyLimitations}[Limited scope and search strategy]
The search string and the Scopus-focused analysis reduce the scope of the research
\end{keyLimitations}
Our study is the first step in a more in-depth analysis. As highlighted in Section \ref{sec:BG}, ESE is a vast field with many empirical study methodologies. We narrowed the selection to case studies and general empirical studies for a pilot study. Therefore, relying heavily on a single search engine, i.e., Scopus, coupled with the limited search string, may not represent the broader landscape of ESE and may present biased observed trends, ultimately affecting the generalization of the results. We address this limitation in the upcoming study.

\begin{keyLimitations}[Expert Selection Bias]
The expert sample is small, potentially hindering generalizability. 
\end{keyLimitations}

The experts were carefully chosen from the leading subfields of Empirical Software Engineering (ESE), each with over 20 years of experience. Many have also served extensively on the editorial boards of top journals and as committee members for flagship conferences. Consequently, while the sample size is small, it is representative of the field’s top researchers.

\begin{keyLimitations}[Information Leakage with Open Access Publications]
LLMs and generative AI are trained on large datasets, including open-access research publications.\end{keyLimitations}

 LLMs are good at providing insights and recommendations but run a significant risk of \textbf{information leakage}. In our context, information leakage occurs when patterns, practices, or inadequate methodologies embedded in the training data are propagated via LLM suggestions. This creates a vicious circle that eventually perpetuates biases and errors in the literature, further impacting statistical strategies and the rigor of empirical analysis.

%indirect bias

%LLMs are usually trained on large datasets, including academic output. This will enable them to suggest methodologies coherent with such datasets' statistical habits. 
\begin{keyLimitations}[Mirroring Distortion with Open Access Publications]
LLMs may suggest methodologies that align with the statistical practices commonly found in the data they were trained on.
\end{keyLimitations}

On the one hand, LLMs' capabilities can facilitate methodological choice, but on the other, they may reinforce methodological flaws implicit in the training data. More precisely, if the training corpus contains statistical misuse, such as performing a parametric test without verifying data distribution, the LLMs may become a vector of diffusion for such bad practices by suggesting them to the user.

For example, a model trained on ESE studies may learn some common but weak hypothesis testing patterns that don't consider data normality assumptions. It thus suggests that such a pattern is an appropriate approach and reinforces its inappropriate use in new studies. This process, therefore, requires careful assessment of an LLM-based recommender system to avoid perpetuating methodological flaws.

\begin{keyLimitations}[The Bias of the ``Hot Topic'']
Published papers often reflect fleeting trends rather than best practices, leading LLMs to recommend widely used but suboptimal methods that reinforce systemic flaws and hinder innovation.
\end{keyLimitations}
Published papers often represent popular, at times transient, trends in which ``methodological flavors [techniques] of the month'' are more common than long-standing best practices. This reliance on prevailing norms can lead LLMs to recommend solutions that, while widely adopted, may be suboptimal and rooted in outdated or flawed approaches. This will consequently drive AI-generated recommendations to reinforce systemic weaknesses in statistical analysis rather than encouraging true methodological innovation, creating a cycle of reliance upon methods that may not serve the evolving needs of empirical research.

%peer review impact
\begin{keyLimitations}[The Issue of Reinventing the Wheel]
Research may trust an LLM-based recommender system that may hallucinate, create new methodological strategies, and strain the peer-review process, which must discern robust methodologies from AI-generated suggestions lacking rigorous justification.
\end{keyLimitations}
This is a dual threat: first, to researchers, by uncritically adopting AI-suggested strategies that might affect the validity and reliability of their findings; second, to the peer-review process, in which reviewers have to make a distinction between genuinely robust methodologies and those that are AI-generated but lack rigorous justification. As noted in prior sections of this study, \textbf{expert evaluations are already concerned with accuracy}. Introducing AI-generated methodologies \textbf{adds another layer of complexity}, further straining the peer-review process. In turn, this limitation is similar to detecting AI-generated text; therefore, it paves the way for models fine-tuned to detect AI-generated methodology.

To address the risk of information leakage and the indirect proliferation of flawed statistical strategies via generative AI, LLMs must be transparent \cite{esposito2024large}. Researchers should record how AI tools helped reviewers and colleagues formulate critical questioning approaches. Besides that, the ESE community should focus on training in GenAI-aware methodologies that would enable researchers to critically evaluate and validate suggestions provided by GenAI rather than uncritically accepting those suggestions. 

In turn, such threats call for curated high-quality datasets for training LLMs based on established best practices in statistical analysis, reducing the risk of perpetuating flawed practices.

Embedding automatic methodological checks into the AI tools may flag potential problems even before researchers can use them, such as preselected mismatches of statistical tests or violations of assumptions about data. Future works will focus on analyzing and reducing the risk of AI-induced errors in methodology and ensure that the benefits derived from generative AI accrue responsibly and with no compromise on stringency or reliability in empirical studies. 

\section{Conclusions}
\label{sec:conclusions}
In conclusion, our findings reveal a concerning prevalence of methodological issues in empirical software engineering (ESE) research over the past 30 years. Our study acts as a first step towards an expansive evaluation across all digital libraries, moving beyond the Scopus-only selection to capture a more comprehensive and in-depth view of our field to critically rethink and propose reforms of the very foundation of the data analysis in our field. Surprisingly, the results from the workshop further highlight significant flaws in the peer-review process, as experts frequently misidentify appropriate statistical tests, models, and methods. These insights, echoed by recent presentations such as Prof. Sira Vegas’s ESEIW 2024 keynote on methodological challenges in deep learning research and the previous empirical studies from \cite{reyes2018statistical}, underscore a widespread issue within our community. Rooted in our own peer-review experiences and inspired by the legacy of the 2006 Dagstuhl seminar, our study strengthens the observation of the pervasive nature of these issues and calls for a critical re-evaluation of current practices. In our future work, we aim to conduct an extensive survey of the state of the art to identify all methodological issues and their reasoning and motivations. Methodological flaws can invalidate entire results, so our ongoing investigation will ultimately present a comprehensive view, framing the core issues that could challenge the foundations of many historical findings in our field.

\section*{Data Availability Statement} 
We provide our raw data and participant responses in our replication package hosted on Zenodo\footnote{\url{https://doi.org/10.5281/zenodo.14073723}}

\section*{Acknowledgments}
The authors thank the workshop participants for their cooperation and willingness to support the study Prof. Travassos is a CNPq Researcher and a FAPERJ CNE in Brazil.
 
%\printcredits

 \bibliographystyle{spbasic}   
 \bibliography{main}

\begin{thebibliography}{45}
\providecommand{\natexlab}[1]{#1}
\providecommand{\url}[1]{{#1}}
\providecommand{\urlprefix}{URL }
\expandafter\ifx\csname urlstyle\endcsname\relax
  \providecommand{\doi}[1]{DOI~\discretionary{}{}{}#1}\else
  \providecommand{\doi}{DOI~\discretionary{}{}{}\begingroup \urlstyle{rm}\Url}\fi
\providecommand{\eprint}[2][]{\url{#2}}

\bibitem[{ACM(2021)}]{ACM_Human_Participants_2021}
ACM (2021) Acm publications policy on research involving human participants and subjects. \url{https://www.acm.org/publications/policies/research-involving-human-participants-and-subjects}

\bibitem[{Agresti(2015)}]{agresti2015foundations}
Agresti A (2015) Foundations of linear and generalized linear models. John Wiley \& Sons

\bibitem[{Armstrong and Hilton(2010)}]{armstrong2010post}
Armstrong RA, Hilton AC (2010) Post hoc tests. Statistical analysis in microbiology: Statnotes pp 39--44

\bibitem[{Basili et~al.(1994)Basili, Caldiera, and Rombach}]{Basili1994}
Basili VR, Caldiera G, Rombach HD (1994) The goal question metric approach. Encyclopedia of Software Engineering

\bibitem[{Basili et~al.(2007)Basili, Rombach, Schneider, Kitchenham, Pfahl, and Selby}]{DBLP:conf/dagstuhl/2006esei}
Basili VR, Rombach HD, Schneider K, Kitchenham BA, Pfahl D, Selby RW (eds) (2007) Empirical Software Engineering Issues. Critical Assessment and Future Directions, International Workshop, Dagstuhl Castle, Germany, June 26-30, 2006. Revised Papers, Lecture Notes in Computer Science, vol 4336, Springer, \doi{10.1007/978-3-540-71301-2}, \urlprefix\url{https://doi.org/10.1007/978-3-540-71301-2}

\bibitem[{Botchkarev(2018)}]{botchkarev2018performance}
Botchkarev A (2018) Performance metrics (error measures) in machine learning regression, forecasting and prognostics: Properties and typology. arXiv preprint arXiv:180903006

\bibitem[{Brown and et~al.(2020)}]{brown2020language}
Brown TB, et~al (2020) Language models are few-shot learners. arXiv preprint arXiv:200514165

\bibitem[{{\c{C}}arka et~al.(2022){\c{C}}arka, Esposito, and Falessi}]{ccarka2022effort}
{\c{C}}arka J, Esposito M, Falessi D (2022) On effort-aware metrics for defect prediction. Empirical Software Engineering 27(6):152

\bibitem[{Esposito and Falessi(2024)}]{esposito2024validate}
Esposito M, Falessi D (2024) Validate: A deep dive into vulnerability prediction datasets. Information and Software Technology p 107448

\bibitem[{Esposito and Palagiano(2024)}]{esposito2024leveraging}
Esposito M, Palagiano F (2024) Leveraging large language models for preliminary security risk analysis: A mission-critical case study. In: Proceedings of the 28th International Conference on Evaluation and Assessment in Software Engineering, ACM, pp 442--445

\bibitem[{Esposito et~al.(2024{\natexlab{a}})Esposito, Falaschi, and Falessi}]{esposito2024extensive}
Esposito M, Falaschi V, Falessi D (2024{\natexlab{a}}) An extensive comparison of static application security testing tools. In: Proceedings of the 28th International Conference on Evaluation and Assessment in Software Engineering, ACM, pp 69--78

\bibitem[{Esposito et~al.(2024{\natexlab{b}})Esposito, Palagiano, Lenarduzzi, and Taibi}]{esposito2024large}
Esposito M, Palagiano F, Lenarduzzi V, Taibi D (2024{\natexlab{b}}) On large language models in mission-critical it governance: Are we ready yet? arXiv preprint arXiv:241211698

\bibitem[{Falessi et~al.(2023)Falessi, Laureani, {\c{C}}arka, Esposito, and Costa}]{falessi2023enhancing}
Falessi D, Laureani SM, {\c{C}}arka J, Esposito M, Costa DAd (2023) Enhancing the defectiveness prediction of methods and classes via jit. Empirical Software Engineering 28(2):37

\bibitem[{Felderer and Travassos(2020)}]{Felderer2020}
Felderer M, Travassos GH (2020) The Evolution of Empirical Methods in Software Engineering, Springer International Publishing, Cham, pp 1--24. \doi{10.1007/978-3-030-32489-6_1}, \urlprefix\url{https://doi.org/10.1007/978-3-030-32489-6_1}

\bibitem[{{Hack Reactor}(2020)}]{hackreactor_history_2020}
{Hack Reactor} (2020) The history of coding and software engineering. \urlprefix\url{https://www.hackreactor.com/resources/the-history-of-coding-and-software-engineering/}, accessed: 2021-05-06

\bibitem[{Hamilton(2018)}]{hamilton2018icse}
Hamilton M (2018) Icse 2018 - plenary sessions - margaret hamilton. YouTube, \urlprefix\url{https://youtu.be/ZbVOF0Uk5lU?si=QrIqU-6di8C-IENs}, 2018 International Conference on Software Engineering, celebrating its 40th anniversary and 50 years of Software Engineering

\bibitem[{H{\"o}fer and Tichy(2007)}]{Hofer2007}
H{\"o}fer A, Tichy WF (2007) Status of Empirical Research in Software Engineering, Springer Berlin Heidelberg, Berlin, Heidelberg, pp 10--19. \doi{10.1007/978-3-540-71301-2_3}, \urlprefix\url{https://doi.org/10.1007/978-3-540-71301-2_3}

\bibitem[{Kaliyadan and Kulkarni(2019)}]{kaliyadan2019types}
Kaliyadan F, Kulkarni V (2019) Types of variables, descriptive statistics, and sample size. Indian dermatology online journal 10(1):82--86

\bibitem[{Kennedy and Sankey(1898)}]{doi:10.1680/imotp.1898.19100}
Kennedy ABW, Sankey HR (1898) The thermal efficiency of steam engines. report of the committee appointed to the council upon the subject of the definition of a standard or standards of thermal efficiency for steam engines: With an introductory note. (including appendixes and plate at back of volume). Minutes of the Proceedings of the Institution of Civil Engineers 134(1898):278--312, \doi{10.1680/imotp.1898.19100}, \urlprefix\url{https://doi.org/10.1680/imotp.1898.19100}, \eprint{https://doi.org/10.1680/imotp.1898.19100}

\bibitem[{Kitchenham(2007)}]{Kitchenham2007}
Kitchenham B (2007) Empirical Paradigm -- The Role of Experiments, Springer Berlin Heidelberg, Berlin, Heidelberg, pp 25--32. \doi{10.1007/978-3-540-71301-2_9}, \urlprefix\url{https://doi.org/10.1007/978-3-540-71301-2_9}

\bibitem[{Kitchenham et~al.(2004)Kitchenham, Dyb{\aa} et~al.}]{DBLP:conf/icse/KitchenhamDJ04}
Kitchenham BA, Dyb{\aa} T, et~al. (2004) Evidence-based software engineering. In: Finkelstein A, Estublier J, Rosenblum DS (eds) 26th International Conference on Software Engineering {(ICSE} 2004), 23-28 May 2004, Edinburgh, United Kingdom, {IEEE} Computer Society, pp 273--281, \doi{10.1109/ICSE.2004.1317449}

\bibitem[{Lee et~al.(2023)Lee, Hartmann, Park, Papailiopoulos, and Lee}]{lee2023prompted}
Lee G, Hartmann V, Park J, Papailiopoulos D, Lee K (2023) Prompted llms as chatbot modules for long open-domain conversation. arXiv preprint arXiv:230504533

\bibitem[{Lenarduzzi et~al.(2017)Lenarduzzi, Stan, Taibi, Venters, and Windegger}]{lenarduzzi2017prioritizing}
Lenarduzzi V, Stan AC, Taibi D, Venters G, Windegger M (2017) Prioritizing corrective maintenance activities for android applications: An industrial case study on android crash reports. In: International Conference on Software Quality, Springer, pp 133--143

\bibitem[{Livingstone(2009)}]{livingstone2009practical}
Livingstone DJ (2009) A practical guide to scientific data analysis. John Wiley \& Sons

\bibitem[{Mahoney(1990)}]{mahoney1990roots}
Mahoney MS (1990) The roots of software engineering. CWI Quarterly 3(4):325--334

\bibitem[{Meloun and Militky(2011)}]{meloun2011statistical}
Meloun M, Militky J (2011) Statistical data analysis: A practical guide. Woodhead Publishing, Limited

\bibitem[{Meyer(2013)}]{meyer2013origin}
Meyer B (2013) The origin of 'software engineering'. \url{https://bertrandmeyer.com/2013/04/04/the-origin-of-software-engineering/}, accessed: 2023-10-21

\bibitem[{{Object Management Group}(2011)}]{bpmn2_specification}
{Object Management Group} (2011) Business Process Model and Notation (BPMN) Version 2.0. \urlprefix\url{https://www.omg.org/spec/BPMN/2.0}, official Specification

\bibitem[{Oivo(2007)}]{Oivo2007}
Oivo M (2007) New Opportunities for Empirical Research, Springer Berlin Heidelberg, Berlin, Heidelberg, pp 22--22. \doi{10.1007/978-3-540-71301-2_6}, \urlprefix\url{https://doi.org/10.1007/978-3-540-71301-2_6}

\bibitem[{Rayl(2008)}]{rayl2008nasa}
Rayl AJS (2008) {NASA Engineers and Scientists—Transforming Dreams Into Reality}. \url{https://www.nasa.gov/50th/50th_magazine/benefits.html}, nASA 50th Anniversary website. Archived from the original on 2010-06-29. Retrieved 2016-11-25

\bibitem[{Reyes et~al.(2018)Reyes, Dieste, Fonseca, and Juristo}]{reyes2018statistical}
Reyes RP, Dieste O, Fonseca ER, Juristo N (2018) Statistical errors in software engineering experiments: a preliminary literature review. In: Proceedings of the 40th International Conference on Software Engineering, Association for Computing Machinery, New York, NY, USA, ICSE '18, p 1195–1206, \doi{10.1145/3180155.3180161}, \urlprefix\url{https://doi.org/10.1145/3180155.3180161}

\bibitem[{Runeson and H{\"{o}}st(2009)}]{DBLP:journals/ese/RunesonH09}
Runeson P, H{\"{o}}st M (2009) Guidelines for conducting and reporting case study research in software engineering. Empir Softw Eng 14(2):131--164

\bibitem[{da~Silva et~al.(2013)da~Silva, Cruz, Gouveia, and Capretz}]{silva2013using}
da~Silva FQB, Cruz SSJO, Gouveia TB, Capretz LF (2013) Using meta-ethnography to synthesize research: {A} worked example of the relations between personality and software team processes. In: 2013 {ACM} / {IEEE} International Symposium on Empirical Software Engineering and Measurement, Baltimore, Maryland, USA, October 10-11, 2013, {IEEE} Computer Society, pp 153--162, \doi{10.1109/ESEM.2013.11}, \urlprefix\url{https://doi.org/10.1109/ESEM.2013.11}

\bibitem[{Sj{\o}berg(2007)}]{Sjoberg2007}
Sj{\o}berg DIK (2007) Knowledge Acquisition in Software Engineering Requires Sharing of Data and Artifacts, Springer Berlin Heidelberg, Berlin, Heidelberg, pp 77--82. \doi{10.1007/978-3-540-71301-2_23}, \urlprefix\url{https://doi.org/10.1007/978-3-540-71301-2_23}

\bibitem[{Soetewey(2021)}]{statsandrWhatStatistical}
Soetewey A (2021) {W}hat statistical test should {I} do? --- statsandr.com. \url{https://statsandr.com/blog/what-statistical-test-should-i-do/}, [Accessed 08-11-2024]

\bibitem[{Staron(2020)}]{staron2020action}
Staron M (2020) Action Research in Software Engineering - Theory and Applications. Springer, \doi{10.1007/978-3-030-32610-4}, \urlprefix\url{https://doi.org/10.1007/978-3-030-32610-4}

\bibitem[{Tedre(2014)}]{tedre2014science}
Tedre M (2014) The science of computing: shaping a discipline. CRC Press

\bibitem[{Torchiano et~al.(2017)Torchiano, Fernández, Travassos, and de~Mello}]{torchiano2017lessons}
Torchiano M, Fernández DM, Travassos GH, de~Mello RM (2017) Lessons learnt in conducting survey research. \urlprefix\url{https://arxiv.org/abs/1702.05744}, \eprint{1702.05744}

\bibitem[{Vegas(2024)}]{Vegas2024}
Vegas S (2024) The method behind the magic: Ensuring reliability in software engineering empirical results. Keynote presentation at the 18th ACM/IEEE International Symposium on Empirical Software Engineering and Measurement (ESEM), \urlprefix\url{https://conf.researchr.org/details/esem-2024/esem-2024-eseiw-esem/1/The-Method-Behind-the-Magic-Ensuring-Reliability-in-Software-Engineering-Empirical-R}

\bibitem[{Vegas and Basili(2007)}]{Vegas2007}
Vegas S, Basili V (2007) Measurement and Model Building Discussion and Summary, Springer Berlin Heidelberg, Berlin, Heidelberg, pp 115--120. \doi{10.1007/978-3-540-71301-2_32}, \urlprefix\url{https://doi.org/10.1007/978-3-540-71301-2_32}

\bibitem[{Vegas and Elbaum(2023)}]{vegas2023pitfalls}
Vegas S, Elbaum S (2023) Pitfalls in experiments with dnn4se: An analysis of the state of the practice. In: Proceedings of the 31st ACM Joint European Software Engineering Conference and Symposium on the Foundations of Software Engineering, Association for Computing Machinery, New York, NY, USA, ESEC/FSE 2023, p 528–540, \doi{10.1145/3611643.3616320}, \urlprefix\url{https://doi.org/10.1145/3611643.3616320}

\bibitem[{Wang et~al.(2023)Wang, Xu, Lan, Hu, Lan, Lee, and Lim}]{wang2023plan}
Wang L, Xu W, Lan Y, Hu Z, Lan Y, Lee RKW, Lim EP (2023) Plan-and-solve prompting: Improving zero-shot chain-of-thought reasoning by large language models. arXiv preprint arXiv:230504091

\bibitem[{White et~al.(2023)White, Fu, Hays, Sandborn, Olea, Gilbert, Elnashar, Spencer-Smith, and Schmidt}]{white2023prompt}
White J, Fu Q, Hays S, Sandborn M, Olea C, Gilbert H, Elnashar A, Spencer-Smith J, Schmidt DC (2023) A prompt pattern catalog to enhance prompt engineering with chatgpt. arXiv preprint arXiv:230211382

\bibitem[{W{\"{o}}hlin et~al.(2012)W{\"{o}}hlin, Runeson, H{\"{o}}st et~al.}]{wholin2012experimentation}
W{\"{o}}hlin C, Runeson P, H{\"{o}}st M, et~al. (2012) Experimentation in Software Engineering. Springer

\bibitem[{Xia et~al.(2020)Xia, Zhang, Zhang, Liang, Peng, and Philip}]{xia2020low}
Xia C, Zhang C, Zhang J, Liang T, Peng H, Philip SY (2020) Low-shot learning in natural language processing. In: 2020 IEEE Second International Conference on Cognitive Machine Intelligence (CogMI), IEEE, pp 185--189

\end{thebibliography}

\end{document}